\documentclass[]{fairmeta}

\usepackage{amsthm,amsmath,amssymb,bm}
\usepackage{graphicx}
\usepackage{algorithm,algorithmic}
\usepackage[numbers,sort&compress]{natbib}
\usepackage[hidelinks,breaklinks]{hyperref}

\newcommand{\A}{\ensuremath{\mathrm{A}}}
\newcommand{\B}{\ensuremath{\mathrm{B}}}
\newcommand{\C}{\ensuremath{\mathrm{C}}}
\newcommand{\D}{\ensuremath{\mathrm{D}}}
\renewcommand{\H}{\ensuremath{\mathrm{H}}}
\newcommand{\T}{\ensuremath{\mathrm{T}}}

\newtheorem{theorem}{Theorem}
\newtheorem{lemma}{Lemma}
\newtheorem{proposition}{Proposition}
\newtheorem{mainproposition}{Proposition}

\title{Exploration stabilizes learning dynamics and promotes cooperation in stochastic games}
\author[1, 2\dagger]{Yuxin Geng}
\author[1, 3\dagger]{Xingru Chen}
\author[4, 5, 6]{Wolfram Barfuss}
\author[1, 2, 7]{Longzhao Liu}
\author[1, 2, 7]{Xin Wang}
\author[1, 8, 9*]{Shaoting Tang}
\author[10, 11*]{Feng Fu}

\affiliation[1]{School of Artificial Intelligence, Beihang University, Beijing 100191, China}
\affiliation[2]{Zhongguancun Laboratory, Beijing 100094, China}
\affiliation[3]{Key Laboratory of Mathematics, Informatics and Behavioral Semantics (Beihang University), Ministry of Education, Beijing 100191, China}
\affiliation[4]{Transdisciplinary Research Area Sustainable Futures, University of Bonn, Germany}
\affiliation[5]{Center for Development Research, University of Bonn, Germany}
\affiliation[6]{Institute for Food and Resource Economics, University of Bonn, Germany}
\affiliation[7]{Beijing Advanced Innovation Center for Future Blockchain and Privacy Computing, Beihang University, Beijing 100191, China}
\affiliation[8]{Institute of Trustworthy Artificial Intelligence, Zhejiang Normal University, Hangzhou 310012, China}
\affiliation[9]{Shandong Key Lab of Complex Medical Intelligence and Aging, Shandong Medical and Pharmaceutical University, Yantai 264003, China}
\affiliation[10]{Department of Mathematics, Dartmouth College, Hanover, NH 03755, USA}
\affiliation[11]{Department of Biomedical Data Science, Geisel School of Medicine at Dartmouth, Lebanon, NH 03756, USA}
\contribution[\dagger]{These authors contributed equally to this work.}

\contribution[*]{Corresponding authors. Emails: \email{tangshaoting@buaa.edu.cn}, \email{feng.fu@dartmouth.edu}}

\begin{document}

\abstract{
Understanding and stabilizing collective learning dynamics is a central challenge in stochastic games. Although exploration is widely used in learning algorithms, its effects on the coupled dynamics of agents and their environment remain insufficiently understood. In this work, we derive analytical equations for entropy-regularized actor--critic learning in stochastic games by separating the timescales of value estimation and policy adaptation. Without exploration, the system preserves a weighted volume in the policy phase space, ruling out asymptotically stable fully mixed equilibria. We provide a unified characterization of exploration's effects on objective curvature, preference dynamics in log-ratio coordinates, and the forgetting of past action preferences. Exploration reduces local volume expansion and can stabilize learning dynamics. In a two-state Matching Pennies game, it suppresses chaotic learning and promotes convergence to a mixed equilibrium. In a Prisoner's Dilemma with environmental feedback, intermediate exploration eliminates the defective attractor and expands the basin of attraction of the cooperative equilibrium. Our findings reveal how exploration shapes collective behavior, informing the design of cooperative and resilient multi-agent systems.
}
\metadata[Keywords]{stochastic games; exploration; entropy regularization; cooperation; multi-agent reinforcement learning}

\maketitle

\clearpage

\section{Introduction}

Learning and adaptation are central to intelligence in both biological and artificial systems~\cite{russell2010artificial}. Reinforcement learning (RL) provides a general framework for understanding how individuals improve their decisions in dynamic and uncertain environments~\cite{sutton1998reinforcement}. Its effectiveness has been demonstrated across diverse fields, including neuroscience, sociology, physics, and artificial intelligence~\cite{fawzi2022discovering, lee2004reinforcement, sgroi2021reinforcement, silver2017mastering, ouyang2022training, geng2026learning}. By learning directly from experience, RL agents can adapt their behavior without a complete model of their surroundings. Temporal-difference (TD) learning is a foundational RL method that updates value estimates based on the discrepancy between the current prediction and a target combining observed rewards with estimated future returns~\cite{sutton1988learning}. These updates are performed in a bootstrapped manner and underpin widely used algorithms, including Q-learning and actor--critic (AC) methods~\cite{watkins1992q, konda1999actor}.

Extending RL to interacting agents gives rise to multi-agent reinforcement learning (MARL), in which agents learn in shared environments and the outcomes of their actions depend on the choices of others. MARL draws on ideas from game theory, control theory, and complex systems science, providing a framework to study learning in both competitive and cooperative settings~\cite{busoniu2008comprehensive, tuyls2005evolutionary, wiering2012reinforcement}. As agents adapt simultaneously, an agent's decision influences not only its own immediate rewards but also environmental transitions and other agents' behavioral adaptation. These systems can exhibit rich emergent collective behaviors, including cooperation and social norms, despite these behaviors not being explicitly encoded in the game's rules or reward structures~\cite{leibo2017multi, jaques2019social, foerster2016learning, ndousse2021emergent}. On the flip side, this coupling also introduces inherent non-stationarity, in the sense that the distributions governing state transitions and rewards from each agent's perspective shift as other agents refine their behaviors~\cite{yang2020overview}. As a result, classical convergence guarantees from single-agent RL do not generally carry over to MARL, and the system can exhibit instability, oscillation, or even collapse during training. 

In such dynamic environments, exploration encourages agents to try alternative actions beyond those that currently appear most rewarding. In RL, exploration is often promoted by incorporating entropy regularization into the learning objective, which encourages agents to adopt more stochastic policies~\cite{mnih2016asynchronous, haarnoja2018soft}. However, how entropy regularization shapes collective learning dynamics, particularly their stability and the emergence of cooperation, remains insufficiently understood~\cite{ahmed2019understanding, barfuss2025collective}. We study this question through AC learning, an approach with foundations in neuroscience and successful applications in various fields such as robotics and large language model (LLM) training~\cite{schulman2017proximal, haarnoja2018soft, ouyang2022training}. We formalize entropy-regularized AC learning in stochastic games~\cite{shapley1953stochastic}, use time-scale separation to derive ordinary differential equations for policy adaptation coupled through environmental feedback, and analyze them using dynamical systems theory. 

We find that the system preserves a weighted volume in the policy phase space without exploration, which precludes asymptotically stable equilibria in the interior of the policy space. We then provide a unified characterization of the effect of exploration on the curvature of the objective function, the dependence on learning history, and the stability of the learning dynamics. Entropy introduces negative curvature into the local policy learning objective and imposes an increasing marginal penalty on policy concentration. In log-ratio coordinates, this exploration induces a linear restoring force, yielding an exponentially decaying memory of past action-value differences. Exploration reduces the local rate of volume expansion and promotes convergence by counteracting the amplification of small policy deviations induced by value feedback. A fixed point is exponentially stable when the exploration rate exceeds the growth rate of these deviations.

We examine our theoretical results through two representative examples of stochastic games: the constant-sum game of two-state Matching Pennies and a Prisoner's Dilemma with environmental feedback~\cite{hilbe2018evolution}. In the two-state Matching Pennies game, the system without exploration contains no asymptotically stable interior equilibrium. As agents care more about future rewards, the learning trajectories transition from periodic orbits to quasi-periodic motion and eventually to chaos. This chaotic regime introduces inherent unpredictability into the learning dynamics. Entropy regularization suppresses this instability and steers the system toward convergence to an equilibrium. In the Prisoner's Dilemma with environmental feedback, exploration can weaken strong preferences for defection, which helps preserve the prosperous environment and raises the value of cooperation when agents care enough about the future. An intermediate level of exploration can yield a globally attracting cooperative equilibrium. These findings provide theoretical insights into the role of exploration in stabilizing learning in dynamic environments, informing the design of more robust and cooperative multi-agent systems.

\section{Related Work}

In the pioneering work of B{\"o}rgers and Sarin~\cite{borgers1997learning}, the continuous-time limit of cross-learning dynamics was shown to be equivalent to the replicator dynamics (RD). This connection uncovers a shared mathematical foundation between RL and evolutionary game theory (EGT)~\cite{wang2024mathematics, du2014aspiration}. In the language of EGT, individuals with higher fitness tend to leave more offspring~\cite{nowak2006evolutionary}, whereas in RL terminology, agents are more likely to choose actions with higher estimated returns. Later, Tuyls et al. extended this framework to myopic Q-learning in normal-form games~\cite{tuyls2003selection}, and Kianercy and Galstyan examined its dynamics for two-player, two-action normal-form games~\cite{kianercy2012dynamics}. Frequency-adjusted multi-agent Q-learning clarified the connection between Q-learning updates and RD~\cite{kaisers2010frequency}. Leonardos and Piliouras further showed that exploration can alter equilibrium stability and induce bifurcations in this Q-learning model~\cite{leonardos2022exploration}.

The above studies mainly focus on one-shot normal-form games, where actions determine payoffs but do not change the environmental state. Nevertheless, real-world scenarios often involve sequential decision-making and dynamic environments. This creates a more intricate learning problem, requiring agents to balance short-term rewards against long-term objectives~\cite{littman1994markov}. Several models have been proposed to connect multi-agent learning in dynamic environments with EGT, including state-coupled RD~\cite{hennes2009state} and RESQ-learning~\cite{hennes2010resq}. Previous studies have also employed time-scale separation to analyze the coupled learning dynamics in such environments~\cite{barfuss2019deterministic, geng2026emergent}. Building on this approach, we derive a unified account of exploration in stochastic games through two-timescale analysis of entropy-regularized AC learning.

\section{Model}

\begin{figure}[!t]
    \centering
    \includegraphics[width=\textwidth]{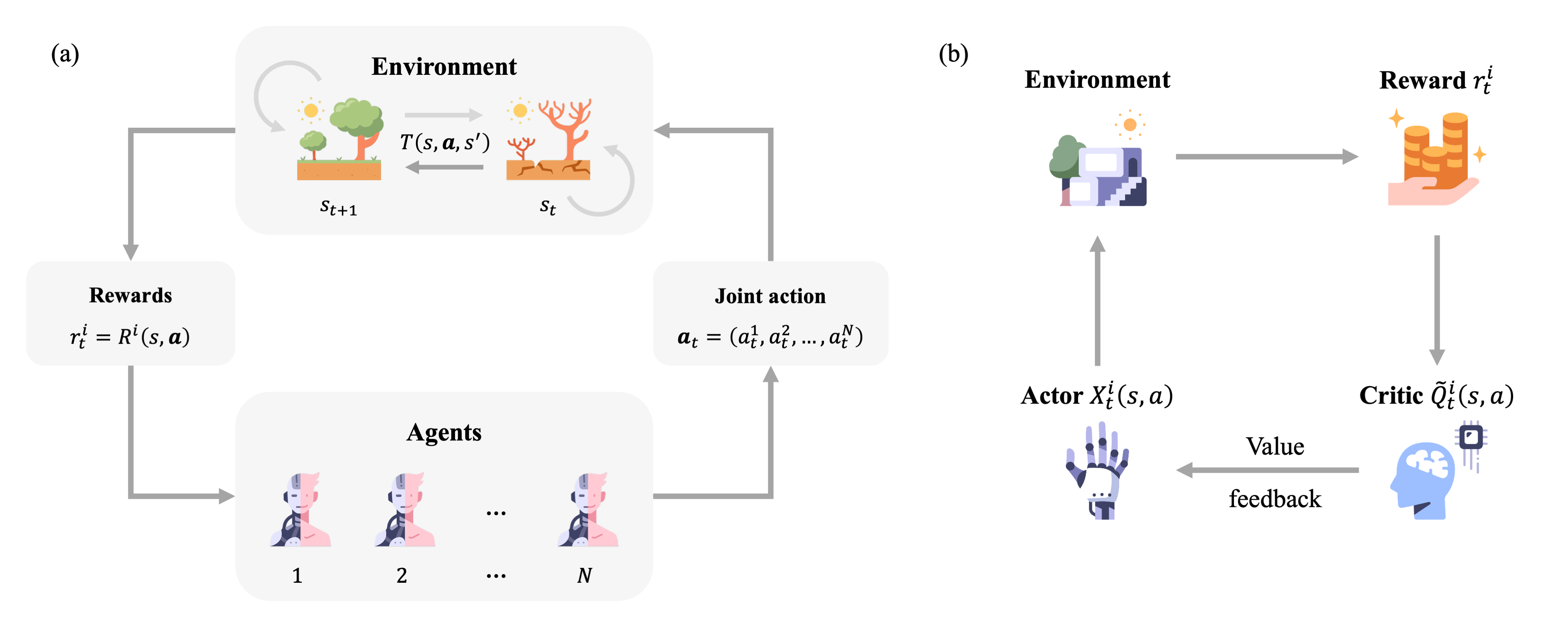}
    \caption{Illustration of multi-agent actor--critic learning in stochastic games. (a) Agents choose actions according to their policies, and their joint actions determine the environmental transitions and each agent's reward. The resulting feedback couples individual learning to collective behavior and future consequences. (b) Each agent adapts its behavior through the actor--critic algorithm. The critic updates action-value estimates using temporal-difference errors computed from observed rewards and value estimates for the subsequent state, according to Eq.~\eqref{eq:critic-update}. The actor uses these estimates to adjust the policy according to Eq.~\eqref{eq:actor-update}.}
    \label{fig:framework}
\end{figure}

We study MARL in dynamic environments modeled as stochastic games, which formally capture both the interactions among agents and the evolution of environmental states. Each agent independently learns via the AC algorithm. Figure~\ref{fig:framework} depicts how agents interact with their environment and update their policies. The main components of our framework are defined in the following subsections, and the corresponding pseudocode is presented in Algorithm~\ref{alg:actor-critic}.

\subsection{Environment}

We consider a stochastic game~\cite{shapley1953stochastic} defined by the tuple $(\mathcal{N}, \mathcal{S}, \mathcal{A}, T, R, \bm{\rho})$. Here, a collection of agents $\mathcal{N} = \{1, 2, \ldots, N\}$ interacts in an environment whose state belongs to $\mathcal{S} = \{s_1, s_2, \ldots, s_K\}$. Each agent chooses an action from the common finite action set $\mathcal A = \{a_1, \ldots, a_M\}$. At time $t$, all agents observe the current state $s_t$ and independently sample their actions from their policies, where $X^i_t(s, a)$ denotes the probability that agent $i$ chooses action $a$ in state $s$. In $s_k$ and $a_m$, the subscripts enumerate states and actions, respectively. The subscript $t$ denotes time, and the superscript $i$ identifies the agent. Agents' joint action $\bm a_t = (a^1_t, \ldots, a^N_t)$ and the current state $s_t$ together determine both the reward $r^i_t = R^i(s_t, \bm a_t)$ received by each agent and the probability $T(s_t, \bm a_t, s_{t + 1})$ of reaching the next state $s_{t + 1}$. Starting from an initial state drawn from a distribution $\bm{\rho}$, these interactions and transitions continue indefinitely.

\subsection{Learning algorithm}

Agents adapt their behavior through AC learning~\cite{konda1999actor}. Each agent maintains its policy $X^i_t(s, a)$ as the actor and an action-value estimate $\tilde Q^i_t(s, a)$ as the critic. The critic estimates the expected discounted cumulative reward obtained by choosing action $a$ in state $s$ and subsequently following the current policies, providing value estimates that guide updates of the actor. At time $t$, each agent $i$ takes an action $a^i_t$ in state $s_t$, receives reward $r^i_t$, and observes the next state $s_{t + 1}$. The interaction sample $(s_t, a^i_t, r^i_t, s_{t + 1})$ is used to update the agent's critic using the TD learning rule~\cite{sutton1998reinforcement}:
\begin{equation}
\tilde Q^i_{t + 1}(s_t, a^i_t)
=
\tilde Q^i_t(s_t, a^i_t)
+ \alpha_Q
\left[
r^i_t + \gamma \tilde V^i_t(s_{t + 1}) - \tilde Q^i_t(s_t, a^i_t)
\right].
    \label{eq:critic-update}
\end{equation}
Here, $\alpha_Q \in (0, 1]$ is the critic learning rate and $\tilde V^i_t(s) = \sum_{a \in \mathcal A}X^i_t(s, a) \tilde Q^i_t(s, a)$ is an estimate of the state value, the expected discounted cumulative reward from state $s$ under the current policies. The update moves the current estimate $\tilde Q^i_t(s_t, a^i_t)$ a fraction $\alpha_Q$ of the way toward the sampled target $r^i_t + \gamma \tilde V^i_t(s_{t + 1})$. The discount factor $\gamma \in [0, 1)$ quantifies the extent to which agents value future rewards, with larger values assigning greater weight to future consequences in the value updates. Only the visited entry $(s_t, a^i_t)$ of the critic is updated at time $t$, and all other entries remain unchanged.

Guided by the critic's value estimates, the actor updates the policy to favor actions with higher values while maintaining a tendency to explore alternatives. We represent this exploration through entropy regularization~\cite{mnih2016asynchronous}. At each state, policy adjustment is guided by the local objective:
\begin{equation}
\mathcal J^i_t(\bm{p}; s)
=
\sum_{a\in\mathcal A}p_a \tilde Q^i_t(s, a) + \eta H(\bm{p}),
\qquad
H(\bm{p}) = -\sum_{a \in \mathcal A}p_a \log p_a,
    \label{eq:local-objective}
\end{equation}
where $\bm{p}$ is a candidate action distribution and $\eta \geq 0$ controls exploration. The first term favors actions with higher estimated returns, while the entropy term encourages agents to retain alternative choices. 

We linearize this objective at the current policy and apply mirror ascent with negative entropy as the mirror function~\cite{beck2003mirror}:
\begin{equation}
    X^i_{t + 1}(s, \cdot)
    = \arg\max_{\bm{p} \in \Delta(\mathcal A)}
    \Bigg\{
        \sum_{a \in \mathcal A}p_a
        \Big[\tilde Q^i_t(s, a) - \eta\log X^i_t(s, a)\Big]
        -\frac{1}{\alpha_X}
        D_{\mathrm{KL}}\bigl(\bm{p} \,\Vert\, X^i_t(s, \cdot)\bigr)
    \Bigg\}.
    \label{eq:mirror-update}
\end{equation}
Here, $\Delta(\mathcal A)$ is the set of action distributions. The term
\begin{equation}
D_{\mathrm{KL}}(\bm{p} \, \Vert \, \bm{x}) = \sum_{a \in \mathcal A} p_a \log(p_a/x_a)
\end{equation} 
is the Kullback--Leibler (KL) divergence between the candidate policy $\bm{p}$ and the current policy $X^i_t(s, \cdot)$. Weighted by $1/\alpha_X$, it controls the magnitude of policy updates by penalizing deviations from the current policy, and $\alpha_X > 0$ is the actor learning rate. The resulting update can be expressed in closed form:
\begin{equation}
X^i_{t + 1}(s,a)
=
\frac{
X^i_t(s, a)
\exp\!\left\{
\alpha_X\left[\tilde Q^i_t(s, a) - \eta \log X^i_t(s, a)\right]
\right\}
}{
\displaystyle\sum_{b \in \mathcal A}
X^i_t(s, b)
\exp\!\left\{
\alpha_X\left[\tilde Q^i_t(s,b) - \eta \log X^i_t(s, b)\right]
\right\}
}.
    \label{eq:actor-update}
\end{equation}
After each interaction, the actor updates its policy at every state using the existing value estimates. We initialize all action probabilities strictly above zero to ensure that this update is well-defined.

\begin{algorithm}
\footnotesize
\caption{Multi-agent actor--critic learning in stochastic games}
\label{alg:actor-critic}
\begin{algorithmic}[1]
    \REQUIRE Stochastic game $(\mathcal N, \mathcal S, \mathcal A, T, R, \bm{\rho})$;
    learning rates $\alpha_Q, \alpha_X$; discount factor $\gamma$; exploration strength $\eta$;
    number of interactions $L$.
    \ENSURE Policies $\bm{X}^i_L$ and value estimates $\tilde{\bm{Q}}^i_L$ for all agents.

    \STATE Initialize policies $X^i_0(s, a)$ and value estimates $\tilde Q^i_0(s, a)$ for all $i,s,a$.
    \STATE Draw the initial state $s_0 \sim \bm{\rho}$.

    \FOR{$t = 0, \ldots, L - 1$}
        \STATE Each agent $i$ independently draws
        $a^i_t \sim X^i_t(s_t, \cdot)$.
        \STATE Execute the joint action
        $\bm a_t = (a^1_t, \ldots, a^N_t)$, and observe
        rewards $r^i_t$ and the next state $s_{t + 1}$ according to $R$ and $T$.

        \FORALL{agents $i\in\mathcal N$}
            \STATE Compute
            $\tilde V^i_t(s_{t + 1}) \gets
            \sum_{a \in \mathcal A}
            X^i_t(s_{t + 1}, a)\tilde Q^i_t(s_{t + 1}, a)$.

            \STATE Copy the current critic: $\tilde{\bm{Q}}^i_{t + 1} \gets \tilde{\bm{Q}}^i_t$.
            \STATE Update the visited entry:
            $\tilde Q^i_{t + 1}(s_t, a^i_t) \gets
            \tilde Q^i_t(s_t, a^i_t) +
            \alpha_Q[r^i_t + \gamma \tilde V^i_t(s_{t + 1})
            -\tilde Q^i_t(s_t, a^i_t)]$.

            \FORALL{states $s \in \mathcal S$}
                \STATE Compute action weights using the old critic:
                $W_t^i(s, a) \gets
                X^i_t(s, a) \exp\!\left\{
                \alpha_X[\tilde Q^i_t(s, a) - \eta \log X^i_t(s, a)]
                \right\}$ for all $a \in \mathcal A$.
                \STATE Normalize to obtain the new policy:
                $X^i_{t + 1}(s, a) \gets
                W_t^i(s, a)/
                \sum_{b \in \mathcal A}W_t^i(s, b)$
                for all $a \in \mathcal A$.
            \ENDFOR
        \ENDFOR
    \ENDFOR
\end{algorithmic}
\end{algorithm}

\section{Results}
\label{sec:results}

\subsection{Dynamics of actor--critic learning in stochastic games}
\label{sec:learning-dynamics}

The coupled adaptation of the actor and critic can lead to inaccurate value estimates, which in turn bias subsequent policy adjustments. Moreover, the learning process contains multiple sources of stochasticity arising from both state transitions and action sampling. To overcome these challenges, we adopt time-scale separation between actor and critic updates, with $\alpha_X \ll \alpha_Q \ll 1$, so that value estimation adapts much faster than policy adaptation~\cite{konda1999actor, barfuss2019deterministic}. Intuitively, the critic can obtain more accurate value estimates of the agents' current behavior before their policies change appreciably. We use a continuous-time description of the learning dynamics, with time measured in units of interactions and updates, and drop the subscript $t$ for clarity. 

For a fixed joint policy profile $\bm X \in [0, 1]^{NKM}$, we first compute agent $i$'s expected rewards and the transition probabilities, given that agent $i$ chooses action $a$ in state $s$:
\begin{equation}
    \begin{aligned}
        R^i(s, a) &= \sum_{\bm a^{-i} \in \mathcal A^{N - 1}}
        \Big[\prod_{j \ne i}X^j(s, a^j)\Big]
        R^i\bigl(s, (a, \bm a^{-i})\bigr),\\
        T^i(s, a, s') &= \sum_{\bm a^{-i} \in \mathcal A^{N - 1}}
        \Big[\prod_{j \ne i}X^j(s, a^j)\Big]
        T\bigl(s, (a, \bm a^{-i}),s' \bigr).
    \end{aligned}
    \label{eq:conditional-environment}
\end{equation}
Here, $\bm a^{-i}$ is the co-players' joint action, and $(a, \bm a^{-i})$ denotes the joint action with agent $i$ choosing $a$. Averaging also over agent $i$'s action yields the state transition probabilities $T(s, s')=\sum_aX^i(s, a)T^i(s, a, s')$ from $s$ to $s'$, which form the matrix $\bm T$. 

For fixed interior policies, we assume that the induced Markov chain of states is irreducible and aperiodic, and mixes sufficiently rapidly relative to the critic updates. Under $\alpha_Q \ll 1$, the chain equilibrates to its stationary distribution $\pi$ and the entry $(s, a)$ is visited with probability $\pi(s)X^i(s, a)$. Averaging the TD error in Eq.~\eqref{eq:critic-update} over these visits, we obtain the continuous-time learning dynamics of the critic:
\begin{equation}
    \frac{\mathrm{d}}{\mathrm{d}t} \tilde Q^i(s, a)
    =\alpha_Q \pi(s) X^i(s, a)
    \left[R^i(s, a) + \gamma \sum_{s' \in \mathcal S}
    T^i(s, a, s') \tilde V^i(s') - \tilde Q^i(s, a)\right].
    \label{eq:critic-dynamics}
\end{equation}
On the fast timescale, the critic converges to the unique Bellman values $Q^i(s, a)$ and $V^i(s)$ satisfying
\begin{equation}
    Q^i(s, a) = R^i(s, a) + \gamma \sum_{s' \in \mathcal S}
    T^i(s, a, s')V^i(s'), \quad \text{and} \quad V^i(s) = \sum_{a \in \mathcal A}X^i(s, a)Q^i(s, a).
    \label{eq:bellman-values}
\end{equation}
On the slow timescale, the actor evolves along the manifold defined by these stationary critic values. We substitute $\bm{Q}^i$ into Eq.~\eqref{eq:actor-update} and expand the update to first order in $\alpha_X$ to obtain the policy dynamics
\begin{equation}
    \frac{\mathrm{d}}{\mathrm{d}t}X^i(s, a)
    = \underbrace{\alpha_X X^i(s, a)
    \left[Q^i(s, a) - V^i(s)\right]}_{\text{exploitation / selection}}
    + \underbrace{\alpha_X \eta X^i(s, a)
    \left[-\log X^i(s, a) - H(X^i(s, \cdot))\right]}_{\text{exploration / mutation}}.
    \label{eq:policy-dynamics}
\end{equation}

This equation separates policy adaptation into a replicator selection term and an entropy-driven exploration term, connecting AC learning to EGT~\cite{nowak2006evolutionary}. In such an interpretation, each agent in each state corresponds to a population, with actions as phenotypes and policy probabilities as their frequencies. The selection term increases the frequency of phenotypes with higher advantage values, analogous to exploitation in RL. The exploration term, on the other hand, plays the role of mutation, preserving population diversity. In the RL interpretation, it encourages agents to sample alternative actions and thereby discover potentially superior strategies.

The effect of exploration becomes simpler when we track log ratios of action probabilities. For $M \geq 2$, we choose $a_M$ as a reference action and define $z^i(s, a) = \log[X^i(s, a)/X^i(s, a_M)]$ for $a \ne a_M$. A positive $z^i(s, a)$ means that an action is favored over the reference action. Taking the difference between the logarithmic derivatives of $X^i(s, a)$ and $X^i(s, a_M)$ in Eq.~\eqref{eq:policy-dynamics} and absorbing $\alpha_X$ into time yields
\begin{equation}
    \frac{\mathrm{d}}{\mathrm{d}t}z^i(s, a)
    = D^i(s, a) - \eta z^i(s, a).
    \label{eq:log-dynamics}
\end{equation}
Here, $D^i(s, a) = Q^i(s, a) - Q^i(s, a_M)$ is the action-value difference relative to the reference action. Equivalently, the equation is expressed in vector form as $\frac{\mathrm{d}}{\mathrm{d}t}\bm{z} = \bm D(\bm{z}) - \eta \bm{z}$. In the following, we use this representation to analyze the learning dynamics of the system and the role of exploration in stabilizing them.

\subsection{Conservation in the absence of exploration}
\label{sec:conservation}

Without exploration ($\eta = 0$), Eq.~\eqref{eq:log-dynamics} reduces to $\dot{\bm{z}} = \bm D(\bm{z})$, so relative action preferences evolve only in response to current value feedback. We find that the dynamics preserve a weighted volume in the phase space, as established by the following theorem.

\begin{theorem}
\label{thm:weighted-conservation}
For $\eta = 0$ and $0 \leq \gamma < 1$, define $w(\bm{z}) = \det(\bm{I} - \gamma \bm T(\bm{z}))$, where $\bm{I}$ is the identity matrix and $\bm T(\bm{z})$ is the transition matrix induced by the current policy. Then
\begin{equation}
    \operatorname{div}\bigl(w(\bm{z}) \bm D(\bm{z})\bigr) = 0.
    \label{eq:weighted-conservation}
\end{equation}
Here, $\operatorname{div}$ denotes divergence with respect to the log-ratio coordinates $\bm{z}$.
Let $d = NK(M - 1)$, and let $\Phi_t$ denote the flow of Eq.~\eqref{eq:log-dynamics}. For any bounded measurable region $U_0 \subset \mathbb R^d$ and its image $U_t = \Phi_t(U_0)$ at $t \geq 0$, we have
\begin{equation}
    \int_{U_t} w(\bm{z}) \,\mathrm{d} \bm{z}
    = \int_{U_0} w(\bm{z}) \,\mathrm{d} \bm{z},
    \qquad t \geq0.
    \label{eq:weighted-volume-integral}
\end{equation}
\end{theorem}
We give the proof of this theorem in Appendix A.3. To interpret Theorem~\ref{thm:weighted-conservation}, imagine a region of initial policies represented in log-ratio coordinates, and track all the trajectories that start within it. The agents' coupled learning dynamics may stretch this region in some directions and compress it in others, but its weighted volume remains unchanged. The weight $w(\bm z)$ offsets changes in ordinary volume caused by feedback from future values. At $\gamma = 0$, $w \equiv 1$, and the theorem reduces to ordinary volume preservation. As shown in Lemma A1 of Appendix A.3, the weight satisfies $0 < (1 - \gamma)^K \leq w(\bm z) \leq (1 + \gamma)^K$ uniformly over all policies. For any bounded measurable initial region $U_0$ of positive volume, the conservation law implies
\begin{equation}
    \int_{U_t} \mathrm d \bm z
    \geq
    \frac{1}{(1 + \gamma)^K}
    \int_{U_t} w(\bm z)\,\mathrm d \bm z
    =
    \frac{1}{(1 + \gamma)^K}
    \int_{U_0} w(\bm z)\,\mathrm d \bm z
    >0.
\end{equation}
This positive lower bound prevents any region of positive volume from contracting into arbitrarily small neighborhoods of an interior equilibrium. Consequently, no equilibrium in the interior of the joint policy simplex is asymptotically stable.

\begin{figure}[!t]
    \centering
    \includegraphics[width=\textwidth]{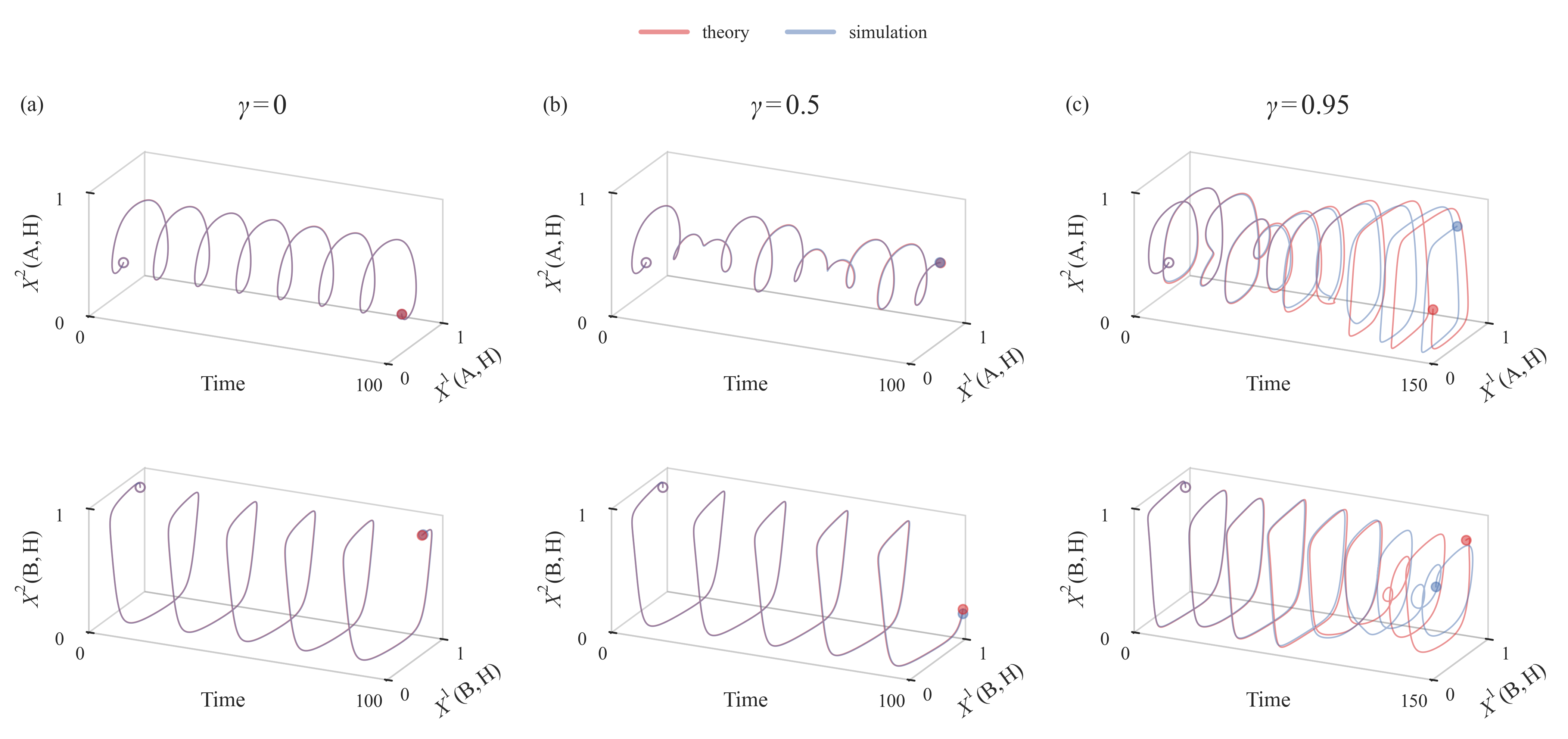}
    \caption{Actor--critic learning without exploration in a two-state Matching Pennies game. Blue curves are agent-based simulations, and red curves are our theoretical predictions, with the trajectories starting from the same initial policy profile. Parameters: $\eta = 0$, $\alpha_Q = 10^{-2}$, and $\alpha_X = 10^{-7}$. The initial policy is $[X^1(\A, \H), X^1(\B, \H), X^2(\A, \H), X^2(\B, \H)] = (0.62, 0.92, 0.23, 0.87)$. }
    \label{fig:mp-chaos}
\end{figure}

As illustrated by the two-state Matching Pennies game in Figure~\ref{fig:mp-chaos}, dynamics that preserve weighted volume can exhibit persistent oscillations and chaos. Boundary attractors also remain possible, as shown by the Prisoner's Dilemma with environmental feedback at $\eta = 0$ in Figure~\ref{fig:donation-branches}(b--c). This is consistent with the conservation law because the boundary lies at infinity in log-ratio coordinates and is not subject to the restriction on interior equilibria. We next examine how exploration through entropy regularization alters this volume-preserving structure and promotes convergence.

\subsection{Dissipation in the presence of exploration}
\label{sec:exploration-stability}

To understand how exploration changes collective learning dynamics, we first examine its effect on individual policy improvement. For fixed action values $Q^i(s, a)$ and $\eta > 0$, entropy introduces strictly negative curvature at an interior policy along every nonzero direction $\bm v$ with $\sum_a v_a = 0$:
\begin{equation}
    \left.
    \frac{\mathrm d^2}{\mathrm d \varepsilon^2}
    \mathcal J\!\left(X^i(s, \cdot) + \varepsilon\bm v\right)
    \right|_{\varepsilon = 0}
    =
    -\eta\sum_a \frac{v_a^2}{X^i(s, a)}
    < 0.
    \label{eq:exploration-entropy-curvature}
\end{equation}
Each additional increase in the probability of a preferred action incurs an increasing penalty as the policy becomes more concentrated. In other words, further concentration requires a larger value advantage to remain favorable. 

This resistance becomes clear when we track the dynamics of policy entropy. Differentiating $H(X^i(s,\cdot))$ along Eq.~\eqref{eq:policy-dynamics} and absorbing $\alpha_X$ into the time variable, we obtain
\begin{equation}
    \frac{\mathrm dH(X^i(s, \cdot))}{\mathrm dt}
    = \underbrace{-\operatorname{Cov}_{a \sim X^i(s, \cdot)}
    \!\left(\log X^i(s, a), Q^i(s, a)\right)}_{\text{exploitation}}
    +\underbrace{\eta \operatorname{Var}_{a \sim X^i(s, \cdot)}
    \!\left(\log X^i(s, a)\right)}_{\text{exploration}}.
    \label{eq:exploration-entropy-balance}
\end{equation}
The covariance term describes whether value feedback reinforces existing preferences. When more probable actions have higher values, exploitation strengthens these preferences and reduces policy entropy, whereas when higher values favor less probable actions, exploitation increases entropy. Exploration contributes a nonnegative term to entropy growth, proportional to the variance of log action probabilities under the current policy.

In log-ratio coordinates, the same marginal entropy penalty becomes the linear damping term $-\eta \bm z$ in Eq.~\eqref{eq:log-dynamics}, which weakens the existing preferences at rate $\eta$. As a result, stronger preferences require larger current value advantages to be maintained. Integrating Eq.~\eqref{eq:log-dynamics} gives
\begin{equation}
    \bm z(t) = e^{-\eta t} \bm z(0)
    +\int_0^t e^{-\eta(t - u)} \bm D(\bm z(u))\,\mathrm du.
    \label{eq:exploration-memory}
\end{equation}
The term $e^{-\eta t} \bm z(0)$ captures the decaying contribution of initial preferences, while the integral accumulates subsequent Bellman action-value differences. Without exploration, all past differences are accumulated in $\bm z(t)$, so an earlier advantage can remain encoded in preferences. Exploration introduces a memory timescale of $1/\eta$: earlier advantages gradually lose their influence, and sustained preferences require consistently higher action values.

The restoring force also reduces the rate of phase-space volume expansion. With $d = NK(M - 1)$ independent log-ratio coordinates, the divergence satisfies
\begin{equation}
    \operatorname{div}(\bm D - \eta \bm z)
    = \operatorname{div}\bm D - d \eta.
    \label{eq:exploration-damping}
\end{equation}
Each coordinate contributes the same decay rate $-\eta$, reducing the local rate of ordinary volume expansion by $d \eta$ at a fixed policy point. 

For $\eta > 0$, at least one interior equilibrium exists (see Proposition A1 of Appendix A.5), characterized by
\begin{equation}
    X^{i\ast}(s, a) = \exp\!\left(Q^i(s, a)/ \eta\right)\Big/\Bigl[\textstyle\sum_{b \in \mathcal A}\exp\!\left(Q^i(s, b)/\eta \right)\Bigr],
    \label{eq:interior-equilibrium}
\end{equation}
where the Bellman values are evaluated at the same joint policy $\bm X^{\ast}$. Equivalently, $\bm D(\bm z^{\ast}) = \eta \bm z^{\ast}$, so current value advantages balance the decay of preferences. Let $\bm{J}^{\ast} = \partial \bm D(\bm z^{\ast})/\partial \bm z$ denote the Jacobian of value feedback at the equilibrium. The following proposition pinpoints the condition for local exponential stability of this equilibrium.

\begin{proposition}
\label{prop:exploration-local-stability}
An interior equilibrium $\bm z^{\ast}$ is exponentially stable if and only if
\begin{equation}
    \max_k \operatorname{Re} \lambda_k(\bm{J}^{\ast}) < \eta,
    \label{eq:exploration-local-stability}
\end{equation}
where $\lambda_k(\bm{J}^{\ast})$ are the eigenvalues of $\bm{J}^{\ast}$.
\end{proposition}
Exploration influences both the location and stability of equilibria. As exploration increases, equilibria shift, and some disappear through bifurcations. The Jacobian of the full dynamics at $\bm z^{\ast}$ is $\bm{J}^{\ast} - \eta \bm{I}$, so exploration shifts each eigenvalue of the Bellman-feedback Jacobian by $-\eta$. The equilibrium is exponentially stable precisely when this restoring effect outweighs the strongest amplification generated by that feedback. Notably, exploration can weaken preferences for defection. Through environmental feedback, this change can eliminate the defective attractor and expand the basin of attraction of cooperation, as depicted in Figure~\ref{fig:donation-branches}(c--e).

\section{Experiments}
\label{sec:experiments}

We now examine how the stabilization mechanism identified above shapes learning in two illustrative stochastic games. In the two-state Matching Pennies game, we show that exploration suppresses chaotic dynamics and promotes convergence to a mixed equilibrium. In the Prisoner's Dilemma with environmental feedback, we examine how exploration interacts with the incentives for cooperation.

\subsection{Two-state Matching Pennies}
\label{sec:mp-experiment}

As a constant-sum example, we consider a two-state Matching Pennies game~\cite{barfuss2019deterministic,cao2019dynamic}. In both states $\A$ and $\B$, each agent can choose between heads ($\H$) and tails ($\T$). We represent the reward structure using the following payoff matrices $\bm{U}(s)$. With agent $1$ as the row player and agent $2$ as the column player, each entry specifies the pair of payoffs received by the agents for a given joint action:
\begin{equation}
\bm{U}(\A) =
\bordermatrix{~ & \H & \T \cr
               \H  & (1, 0) & (0, 1) \cr
               \T  & (0, 1) & (1, 0) \cr},
\quad
\bm{U}(\B) =
\bordermatrix{~ & \H & \T \cr
               \H & (0, 1) & (1, 0) \cr
               \T & (1, 0) & (0, 1) \cr}.
\end{equation}
In state $\A$, matching actions reward agent $1$, whereas mismatched actions reward agent $2$. These roles are reversed in state $\B$. The environment switches to the other state whenever agent $1$ chooses $\H$ and remains in its current state otherwise.

We first set $\eta = 0$ to examine learning dynamics without exploration. Figure~\ref{fig:mp-chaos} compares the agent-based simulations with our theoretical predictions. When $\gamma = 0$, agents respond only to immediate rewards, and the policy dynamics in the two states decouple. Each state exhibits closed cycles in its policy plane. At $\gamma = 0.5$, future rewards couple policy adjustments across the two states. The simulated trajectories become more intricate, but remain close to the theoretical predictions. At $\gamma = 0.95$, the simulated and theoretical trajectories initially agree but subsequently diverge.

\begin{figure}[!t]
    \centering
    \includegraphics[width=\textwidth]{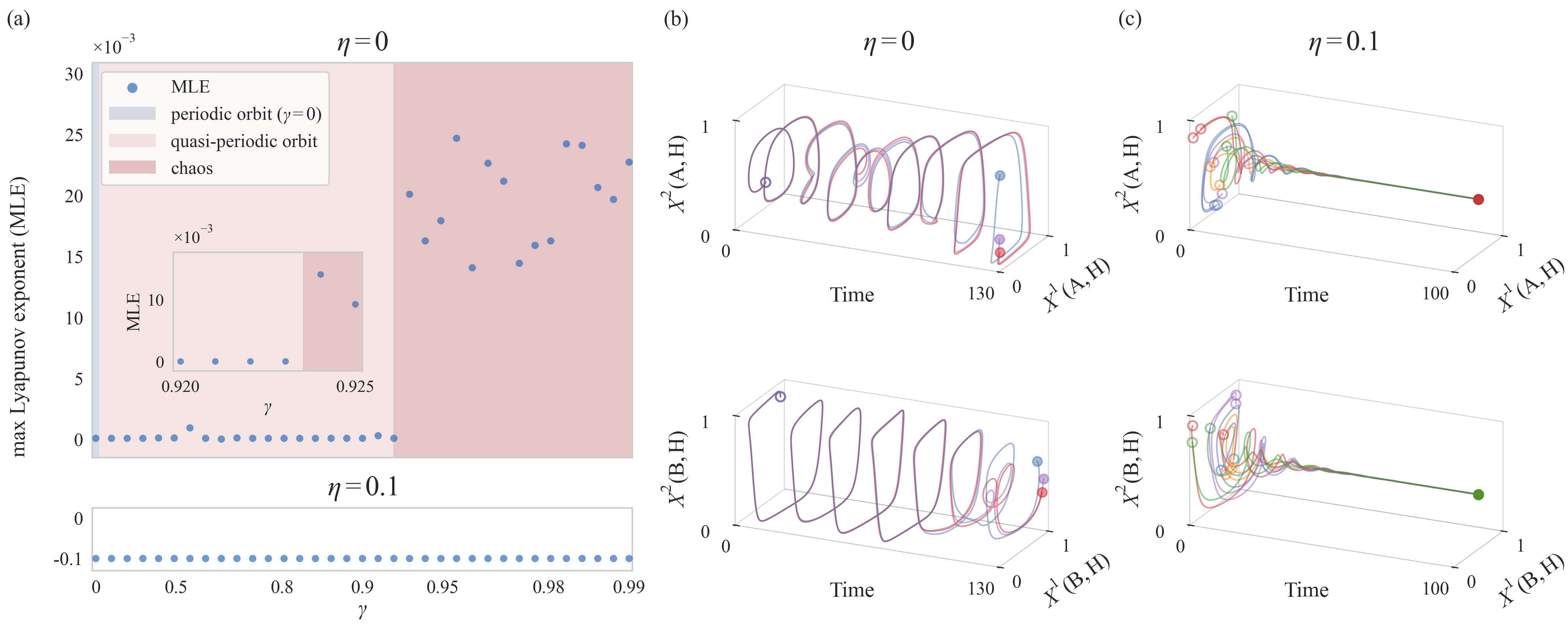}
    \caption{Exploration and learning dynamics in the two-state Matching Pennies game. (a) Maximum Lyapunov exponents at different exploration rates $\eta$. Without exploration, the exponent becomes positive at large discount factors. With exploration, it remains negative across the range, indicating long-term exponential decay of infinitesimal perturbations along these trajectories. (b) Starting from the reference initial policy and two perturbed policies obtained by adding or subtracting $10^{-3}$ from each of its four initial probabilities of action $\H$, three theoretical trajectories at $\eta = 0$ diverge from each other. (c) Starting from independently sampled interior policies, the ten trajectories at $\eta = 0.1$ converge to the mixed equilibrium point. In (b--c), the upper and lower plots show states $\A$ and $\B$, respectively. Open and filled dots indicate initial and final policies. Parameters: (b--c) $\gamma = 0.95$, and the initial policy in (a--b) is $[X^1(\A,\H), X^1(\B,\H), X^2(\A,\H), X^2(\B,\H)] = (0.62, 0.92, 0.23, 0.87)$.}
    \label{fig:mp-exploration}
\end{figure}

These observations illustrate the conservation law in Theorem~\ref{thm:weighted-conservation}, which excludes asymptotically stable interior equilibria while permitting expansion in some directions and contraction in others. To understand the separation between simulated and theoretical trajectories in Figure~\ref{fig:mp-chaos}(c), we examine the sensitivity of the theoretical dynamics to initial conditions while keeping $\eta = 0$. It turns out that small perturbations to the initial policy cause the theoretical trajectories to diverge over time (Figure~\ref{fig:mp-exploration}(b)). This sensitivity can be further quantified by the maximum Lyapunov exponent, a measure of the long-term exponential rate at which infinitesimally close trajectories separate~\cite{souchay2010dynamics}. The exponent is close to zero over much of the discount range and becomes positive at larger discount factors (Figure~\ref{fig:mp-exploration}(a)).

We then introduce exploration with $\eta > 0$ to examine whether it can suppress this chaotic behavior. The corresponding damping counteracts the expansion and thus the separation of nearby trajectories. At $\eta = 0.1$, the maximum Lyapunov exponent is negative across the range of $\gamma$ (Figure~\ref{fig:mp-exploration}(a)). At $\gamma = 0.95$, trajectories from ten sampled interior policies exhibit decaying oscillations and finally converge to the equilibrium $X^i(\A,\H) = X^i(\B,\H) = 1/2$ (Figure~\ref{fig:mp-exploration}(c)). The results demonstrate that exploration can suppress chaotic behavior and stabilize learning.

\subsection{Prisoner's Dilemma with environmental feedback}
\label{sec:donation-experiment}

\begin{figure}[!t]
    \centering
    \includegraphics[width=\textwidth]{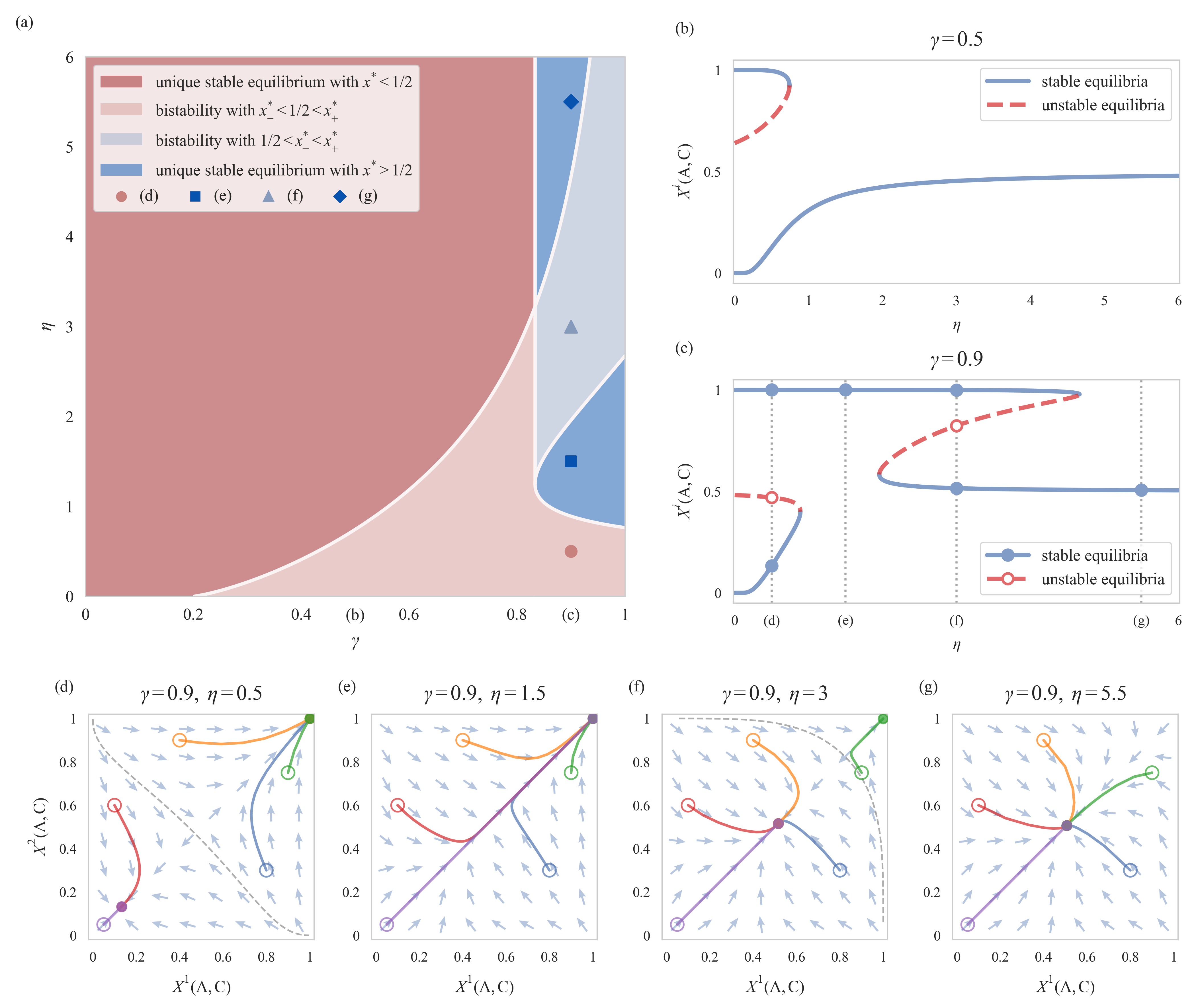}
    \caption{Future rewards and exploration shape cooperation under environmental feedback. (a) Regions in the $(\gamma, \eta)$ plane classified by the number of stable equilibria and whether their cooperation probabilities $x^{\ast}$, $x^{\ast}_-$, and $x^{\ast}_+$ lie above or below $1/2$. (b--c) Cooperation probability $X^{i\ast}(\A, \C)$ at equilibrium as $\eta$ varies at $\gamma = 0.5$ and $0.9$, respectively. Blue solid curves denote stable equilibria; red dashed curves denote unstable equilibria. The two players have identical cooperation probabilities at interior equilibria. (d--g) Learning dynamics at $\gamma = 0.9$ and $\eta = 0.5, 1.5, 3, 5.5$, respectively. The axes are the players' cooperation probabilities in state $\A$. Arrows indicate the direction of policy change; gray dashed curves separate attraction basins. Colored curves show trajectories from the same five initial profiles, with open and filled circles marking initial and final policies. Symbols in (a) and vertical guides in (c) identify the parameter settings for (d--g). Parameters: $b = 5$, $c = 1$, and $q = 0.05$.}
    \label{fig:donation-branches}
\end{figure}

To investigate the effect of exploration on multistability and the emergence of cooperation, we consider the Prisoner's Dilemma with environmental feedback~\cite{hilbe2018evolution}. The game involves two agents and two states. The prosperous state $\A$ is characterized by a Prisoner's Dilemma, where agents can choose between cooperation ($\C$) and defection ($\D$). The payoff matrix is given by
\begin{equation}
    \bm U(\A)=
    \bordermatrix{~ & \C & \D \cr
        \C & (b - c, b - c) & (-c, b) \cr
        \D & (b, -c) & (0, 0) \cr},
    \label{eq:donation-payoffs}
\end{equation}
where $b > c > 0$. Mutual cooperation in $\A$ keeps the environment prosperous, whereas any defection triggers a transition to the degraded state $\B$, where both agents receive zero rewards. From state $\B$, the environment returns to $\A$ with probability $q$, independently of the agents' actions, and otherwise remains in $\B$.

In the normal-form Prisoner's Dilemma, without environmental transitions, defection is the dominant action for both players~\cite{nowak2006evolutionary, barfuss2019deterministic}. Intuitively, cooperation imposes a personal cost $c$ on the player and delivers a benefit $b$ to the co-player, whereas defection incurs no cost and provides no benefit to the co-player. Regardless of what the co-player does, defection always yields a higher immediate payoff. However, mutual cooperation yields a payoff of $b - c$ for each player, which exceeds the zero payoff from mutual defection. The situation changes when state transitions are introduced, as players must then balance short-term gains from defection against the risk of state degradation and the potential loss of long-term collective benefits~\cite{hilbe2018evolution}. 

We show the learning dynamics and equilibria for different exploration rates $\eta$ and discount factors $\gamma$ in Figure~\ref{fig:donation-branches}. Panel (a) classifies regions in the $(\gamma, \eta)$ plane by the number and cooperation levels of stable equilibria, and panels (b) and (c) show equilibrium branches at $\gamma = 0.5$ and $0.9$, respectively. Without exploration, cooperation can be sustained when agents care enough about the future, but this does not ensure its recovery after a collapse. Environmental feedback can stabilize cooperation while leaving defection as an attractor, so perturbations that push policies into the defective basin prevent cooperation from recovering. Exploration can change this dependence on initial behavior by loosening strong preferences for defection. 

At $\gamma = 0.9$, weak exploration preserves the coexistence of cooperative and defective equilibria. As exploration increases, the defection branch approaches the unstable branch until the two merge and disappear, resulting in a globally attracting cooperative equilibrium under intermediate exploration rates (Figure~\ref{fig:donation-branches}(c)). By increasing the frequency of cooperative interactions, exploration helps maintain the prosperous environment, where the future benefits of cooperation further reinforce cooperative behavior.

Because actions in state $\B$ affect neither rewards nor transitions, we focus on the players' cooperation probabilities in state $\A$. The basins of attraction of the equilibria in Figure~\ref{fig:donation-branches}(c) are visualized in Figure~\ref{fig:donation-branches}(d--g) under heterogeneous initial policies. Under weak exploration (panel (d)), the cooperative basin occupies only about half of the policy space, and sufficiently strong initial preferences for defection prevent agents from reaching cooperation. At an intermediate exploration level (panel (e)), this basin expands to all interior initial policies and all trajectories converge to the cooperative equilibrium. However, in panels (f) and (g), the restoring force that weakens preferences for defection also opposes strong cooperative preferences, so excessive exploration can undermine the cooperation sustained by future rewards. 

The above results reveal complementary roles for exploration and environmental transitions. Environmental feedback can sustain cooperation once established, yet agents may remain trapped near defection. Exploration complements the incentives supplied by environmental feedback by helping cooperation recover when those incentives alone cannot overcome the trap of defection.

\section{Discussion}
\label{sec:discussion}

In Eq.~\eqref{eq:interior-equilibrium}, we characterize a form of bounded rationality at equilibrium: agents prefer actions with higher values while retaining alternatives. This flexibility can help agents adapt to changing incentives and promote convergence of the collective learning dynamics. Our results connect objective curvature, policy entropy evolution, and preference dynamics in log-ratio coordinates. We show how a single entropy regularization term influences learning history and learning outcomes. These perspectives complement the role of exploration in injecting stochasticity and acquiring information~\cite{sutton1998reinforcement, ahmed2019understanding}.

Exploration changes how value differences translate into behavioral adaptation, even when agents accurately evaluate their current policies. The central mechanism is that maintaining a strong preference requires the favored action to retain a higher value. This matters especially in stochastic game environments, where different agents' behavior and the environment coevolve. In particular, an agent's policy update changes the induced transition probabilities and expected rewards of other agents, which in turn alters the value estimates steering subsequent updates. By diminishing the influence of accumulated historical preferences on current decisions, exploration can reduce the sensitivity of learning dynamics to initial conditions.

In social dilemmas, exploration does not itself confer a payoff advantage on cooperation, but keeps cooperative actions available to be tried. Mechanisms such as direct reciprocity~\cite{chen2023outlearning, glynatsi2024conditional,pal2026large}, indirect reciprocity~\cite{nowak1998evolution, ohtsuki2006leading, santos2018social}, and environmental stochasticity~\cite{hilbe2018evolution,liu2023coevolutionary} can provide incentives for cooperation through repeated interactions, reputation, or environmental transitions~\cite{nowak2006five,geng2026learning}. Our example with environmental feedback demonstrates how exploration can help cooperation recover by eliminating the defective attractor and expanding the cooperative basin (Figure~\ref{fig:donation-branches}). More generally, when an existing mechanism can sustain cooperation, exploration may help expand its basin of attraction and enhance its resilience.

Our analysis also provides a theoretical perspective for studying other types of policy regularization and structured exploration. Up to an additive constant, the entropy regularization we considered is equivalent to penalizing the KL divergence from a uniform policy, which can be replaced by other reference policies. This offers a dynamical perspective on KL constraints in LLM post-training, where the reference may be a supervised fine-tuned model~\cite{ouyang2022training}. The same regularization term that limits departures from the reference policy also determines the decay of accumulated preferences. 

The effects of exploration also depend on the level at which behavior is represented. In our model, exploration operates over a given set of primitive actions. For complex tasks, a successful behavior often requires coordinated sequences of actions. For example, an LLM needs to generate a sequence of tokens to complete a coding task, while agents in sequential social dilemmas need to coordinate a series of actions to achieve cooperation. Increasing randomness purely in low-level actions may not increase the likelihood of attempting different coherent behaviors and may even disrupt their execution. A promising direction is to extend our analysis to temporally extended actions~\cite{sutton1999between} and examine the joint influence of behavioral abstraction and exploration on the learning dynamics, convergence, and stability. One can also study the synergy between exploration and other mechanisms that promote cooperation~\cite{chen2015competition, zhang2014evolution} and exploration in structured populations~\cite{rong2010feedback, allen2017evolutionary, fu2026optimal}. Together, our findings establish a theoretical basis for understanding exploration and inform the design of cooperative and resilient multi-agent systems.

\section*{Acknowledgements}

This work is supported by the National Cyber Security-National Science and Technology Major Project (2025ZD1503700), the National Natural Science Foundation of China (12526528, 12425114, 12501702, 12671643, 62441617), the Taishan Scholars Program of Shandong Province (tstp20230635), the Key R\&D Program of Shandong Province (2026CXPT283), the Fundamental Research Funds for the Central Universities, the Opening Project of the State Key Laboratory of General Artificial Intelligence (Project NSKLAGI2025OP16), the special funding for the ``Case-by-Case-Introduction of Top Talent (Teams)'', and the Beijing Advanced Innovation Center for Future Blockchain and Privacy Computing.

\clearpage
\appendix
\numberwithin{equation}{section}
\numberwithin{theorem}{section}
\numberwithin{lemma}{section}
\numberwithin{proposition}{section}
\renewcommand{\theequation}{\thesection\arabic{equation}}
\renewcommand{\thetheorem}{\thesection\arabic{theorem}}
\renewcommand{\thelemma}{\thesection\arabic{lemma}}
\renewcommand{\theproposition}{\thesection\arabic{proposition}}

\section{Results}
\label{sec:si-results}
\renewcommand{\theHequation}{\Alph{section}.\arabic{equation}}

\subsection{Dynamics of actor--critic learning}
\label{sec:si-ac-dynamics}

To describe the learning dynamics, we first average the rewards and transitions over the co-players' actions. For a fixed policy profile $\bm X=(\bm{X}^1,\ldots,\bm{X}^N)$, the conditional transition probability $T^i(s, a, s')$ from state $s$ to $s'$ and $i$'s conditional expected reward $R^i(s, a)$ in state $s$, given that player $i$ takes action $a$, are
\begin{align}
    T^i(s,a,s')
    &=\sum_{\bm a^{-i}\in\mathcal A^{N-1}}
      \Big[\prod_{j\ne i}X^j(s,a^j)\Big]
      T\bigl(s,(a,\bm a^{-i}),s'\bigr),
    \label{eq:si-conditional-transition}\\
    R^i(s,a)
    &=\sum_{\bm a^{-i}\in\mathcal A^{N-1}}
      \Big[\prod_{j\ne i}X^j(s,a^j)\Big]
      R^i\bigl(s,(a,\bm a^{-i})\bigr).
    \label{eq:si-conditional-reward}
\end{align}
Here, $(a,\bm a^{-i})$ inserts $a$ in player $i$'s position. The induced unconditional state transition probability $T(s,s')$ from $s$ to $s'$ and $i$'s unconditional expected reward $R^i(s)$ in state $s$ are
\begin{equation}
    T(s,s')=\sum_{a\in\mathcal A}X^i(s,a)T^i(s,a,s'),
    \qquad
    R^i(s)=\sum_{a\in\mathcal A}X^i(s,a)R^i(s,a).
    \label{eq:si-induced-kernel}
\end{equation}
We write $\bm T=[T(s,s')]_{s,s'\in\mathcal S}$ and
$\bm V^i=(V^i(s))_{s\in\mathcal S}$ for the induced transition matrix and value vector.
Their dependence on the current policy profile is left implicit.

We next derive the continuous-time learning dynamics under time-scale separation $\alpha_X\ll\alpha_Q\ll1$~\cite{konda1999actor,barfuss2019deterministic}. Suppose the induced Markov chain of states is irreducible and aperiodic, and let $\pi(s)$ be the corresponding stationary state distribution. The learning dynamics of the critic are then approximated by the ODE
\begin{equation}
    \frac{\mathrm{d}}{\mathrm{d}t}\tilde Q^i(s,a)=\alpha_Q\pi(s)X^i(s,a)
    \left[R^i(s,a)
    +\gamma\sum_{s'\in\mathcal S}T^i(s,a,s')\tilde V^i(s')
    -\tilde Q^i(s,a)\right],
    \label{eq:si-critic-ode}
\end{equation}
where $\tilde V^i(s)=\sum_aX^i(s,a)\tilde Q^i(s,a)$. The factor $\pi(s)X^i(s,a)$ is the probability of visiting the critic entry. On the fast timescale, $\tilde{\bm{Q}}$ evolves much faster than $\bm{X}$ and converges to the unique Bellman values $\bm{Q}^i$ and $\bm{V}^i$ satisfying
\begin{align}
    Q^i(s,a)&=R^i(s,a)
    +\gamma\sum_{s'\in\mathcal S}T^i(s,a,s')V^i(s'),
    \label{eq:si-exact-action-value}\\
    V^i(s)&=\sum_{a\in\mathcal A}X^i(s,a)Q^i(s,a).
    \label{eq:si-exact-state-value}
\end{align}
On the slow timescale, $\bm{X}$ evolves on the stationary manifold defined by these Q-values. Expanding the actor update, Eq.~(5) in the main text, to first order in $\alpha_X$ gives the learning dynamics of the actor:
\begin{equation}
    \frac{\mathrm{d}}{\mathrm{d}t}X^i(s,a)=\alpha_X X^i(s,a)
    \left[Q^i(s,a)-V^i(s)
    -\eta\left(\log X^i(s,a)+H(X^i(s,\cdot))\right)\right].
    \label{eq:si-replicator-ode}
\end{equation}

\subsection{Learning dynamics in log-ratio coordinates}
\label{sec:si-log-dynamics}

Choose $a_M$ as the reference action and define the independent log ratios
\begin{equation}
    z^i(s,a)=\log\frac{X^i(s,a)}{X^i(s,a_M)},
    \qquad a\ne a_M.
    \label{eq:si-log-ratio}
\end{equation}
There are $d=NK(M-1)$ coordinates, where we assume $M\geq2$. Their inverse is
\begin{equation}
    X^i(s,a_M)=\frac{1}{1+\sum_{b\ne a_M}e^{z^i(s,b)}},
    \qquad
    X^i(s,a)=\frac{e^{z^i(s,a)}}{1+\sum_{b\ne a_M}e^{z^i(s,b)}}.
    \label{eq:si-log-inverse}
\end{equation}
Taking action ratios in the actor update, Eq.~(5) in the main text, cancels its normalization and yields the exact discrete identity
\begin{equation}
    z^i_{t+1}(s,a)-z^i_t(s,a)
    =\alpha_X\left[\tilde Q^i_t(s,a)-\tilde Q^i_t(s,a_M)-\eta z^i_t(s,a)\right].
    \label{eq:si-discrete-log-update}
\end{equation}
Measuring time in actor units absorbs the common factor $\alpha_X$. Under time-scale separation, the continuous-time dynamics of $\bm{z}$ are
\begin{equation}
    \frac{\mathrm{d}}{\mathrm{d}t}\bm{z}=\bm D(\bm{z})-\eta \bm{z},
    \qquad
    D^i(s,a)=Q^i(s,a)-Q^i(s,a_M).
    \label{eq:si-reduced-log-ode}
\end{equation}
Here, $\bm D(\bm{z})$ collects the action-value gaps $D^i(s,a)$ evaluated at $\bm X(\bm{z})$. Entropy therefore gives linear damping in these coordinates.

Let $R_*=\max_{i,s,\bm a}|R^i(s,\bm a)|$. The Bellman equations give
\begin{equation}
    \|(\bm{I}-\gamma\bm T)^{-1}\|_\infty=\frac{1}{1-\gamma},
    \qquad
    \|\bm D(\bm{z})\|_\infty\leq B:=\frac{2R_*}{1-\gamma}.
    \label{eq:si-gap-bound}
\end{equation}
The Bellman solution is smooth on the compact product of policy simplices. Together with Eq.~\eqref{eq:si-log-inverse}, this makes $\bm D$ smooth with globally bounded first derivatives. Consequently Eq.~\eqref{eq:si-reduced-log-ode} has a unique solution for every finite initial $\bm{z}$ and all finite times. At $\gamma=0$, the action gaps depend only on policies in the same state; continuation values can couple different states when $\gamma>0$.

\subsection{Conservation in the absence of exploration}
\label{sec:si-invariant-volume}

We first establish that the Bellman determinant defines a positive, uniformly bounded volume weight.

\begin{lemma}
\label{lem:si-volume-weight}
For $\gamma\in[0,1)$, define
\begin{equation}
    w(\bm{z})=\det(\bm{I}-\gamma\bm T(\bm{z})),
    \label{eq:si-volume-density}
\end{equation}
where $\bm T(\bm{z})$ is the transition matrix induced by $\bm X(\bm{z})$. For every $\bm{z}\in\mathbb R^d$,
\begin{equation}
    0<(1-\gamma)^K\leq w(\bm{z})\leq(1+\gamma)^K.
    \label{eq:si-volume-weight-bounds}
\end{equation}
\end{lemma}

\begin{proof}
Fix $\bm{z}$. Since $\bm T(\bm{z})$ is a stochastic matrix, each of its eigenvalues $\lambda_1,\ldots,\lambda_K$ satisfies $|\lambda_k|\leq1$. Hence
\begin{equation*}
    1-\gamma\leq1-\gamma|\lambda_k|
    \leq|1-\gamma\lambda_k|
    \leq1+\gamma|\lambda_k|\leq1+\gamma.
\end{equation*}
In particular, $\bm{I}-\gamma\bm T(\bm{z})$ is nonsingular. Since the determinant is real, continuous in $\gamma$, equal to one at $\gamma=0$, and nonzero for all $\gamma<1$, it remains positive.
Taking the product of the eigenvalue bounds therefore gives
\begin{equation*}
    (1-\gamma)^K
    \leq\prod_{k=1}^K|1-\gamma\lambda_k|
    =|\det(\bm{I}-\gamma\bm T(\bm{z}))|=w(\bm{z})
    \leq(1+\gamma)^K.
\end{equation*}
\end{proof}

The following theorem shows that learning without entropy regularization preserves the volume weighted by $w(\bm{z})$. For an initial region $U_0$ in log-ratio space, let $U_t$ consist of the points reached at time $t$ by trajectories starting in $U_0$.

\begin{theorem}[Bellman-weighted invariant volume]
\label{thm:si-invariant-volume}
For $\eta=0$, the vector field $\bm D$ and the weight in Eq.~\eqref{eq:si-volume-density} satisfy
\begin{equation}
    \operatorname{div}(w\bm D)=0.
    \label{eq:si-invariant-volume}
\end{equation}
Let $\Phi_t$ denote the flow of $\mathrm{d}\bm{z}/\mathrm{d}t=\bm D(\bm{z})$. For any bounded measurable region $U_0\subset\mathbb R^d$ and $U_t=\Phi_t(U_0)$,
\begin{equation}
    \int_{U_t}w(\bm{z})\,\mathrm{d}\bm{z}
    =\int_{U_0}w(\bm{z})\,\mathrm{d}\bm{z},
    \qquad t\geq0.
    \label{eq:si-weighted-volume-integral}
\end{equation}
\end{theorem}

\begin{proof}
Fix a player $i$ and a state $s$. We first differentiate with respect to $z^i(s,\tilde a)$, where $\tilde a\ne a_M$, keeping the state fixed and allowing the action $a$ being differentiated to differ from $\tilde a$. By differentiating Eq.~\eqref{eq:si-log-inverse}, we obtain
\begin{equation}
    \frac{\partial X^i(s,a)}{\partial z^i(s,\tilde a)}
    =X^i(s,a)
    \left[\mathbf1_{\{a=\tilde a\}}-X^i(s,\tilde a)\right],
    \qquad a\in\mathcal A.
    \label{eq:si-policy-derivative}
\end{equation}
Thus only player $i$'s policy at state $s$ changes; the policies of other players have zero derivative with respect to this coordinate.

The conditional quantities $T^i(s,a,s')$ and $R^i(s,a)$ hold player $i$'s action fixed at $a$ and average only over the co-players' actions. They are therefore independent of player $i$'s own policy. Using $T(s,s')=\sum_aX^i(s,a)T^i(s,a,s')$, we obtain
\begin{equation*}
    \begin{aligned}
        \frac{\partial T(s,s')}{\partial z^i(s,\tilde a)}
        &=\sum_aX^i(s,a)
        \left[\mathbf1_{\{a=\tilde a\}}-X^i(s,\tilde a)\right]T^i(s,a,s')\\
        &=X^i(s,\tilde a)
        \left[T^i(s,\tilde a,s')-\sum_aX^i(s,a)T^i(s,a,s')\right]\\
        &=X^i(s,\tilde a)
        \left[T^i(s,\tilde a,s')-T(s,s')\right].
    \end{aligned}
\end{equation*}
For the derivative with respect to a general variable $z^i(\tilde s,\tilde a)$ with $\tilde a\ne a_M$, $T(s,s')$ and $R^i(s)$ depend only on the policy at state $s$. Hence we have
\begin{equation}
    \begin{aligned}
        \frac{\partial T(s,s')}{\partial z^i(\tilde s,\tilde a)}
        &=\mathbf1_{\{s=\tilde s\}}X^i(\tilde s,\tilde a)
        \left[T^i(\tilde s,\tilde a,s')-T(\tilde s,s')\right],\\
        \frac{\partial R^i(s)}{\partial z^i(\tilde s,\tilde a)}
        &=\mathbf1_{\{s=\tilde s\}}X^i(\tilde s,\tilde a)
        \left[R^i(\tilde s,\tilde a)-R^i(\tilde s)\right].
    \end{aligned}
    \label{eq:si-transition-reward-derivatives}
\end{equation}

Next, write the state-value Bellman equation in components as
$V^i(s)=R^i(s)+\gamma\sum_{s'}T(s,s')V^i(s')$.
Both the transition probabilities and the continuation values depend on the policy. Applying the product rule and substituting Eq.~\eqref{eq:si-transition-reward-derivatives} yields
\begin{equation*}
    \begin{aligned}
        \frac{\partial V^i(s)}{\partial z^i(\tilde s,\tilde a)}
        &=\frac{\partial R^i(s)}{\partial z^i(\tilde s,\tilde a)}
        +\gamma\sum_{s'}\frac{\partial T(s,s')}{\partial z^i(\tilde s,\tilde a)}V^i(s')
        +\gamma\sum_{s'}T(s,s')\frac{\partial V^i(s')}{\partial z^i(\tilde s,\tilde a)}\\
        &=\mathbf1_{\{s=\tilde s\}}X^i(\tilde s,\tilde a)
        \left[R^i(\tilde s,\tilde a)-R^i(\tilde s)
        +\gamma\sum_{s'}\bigl(T^i(\tilde s,\tilde a,s')-T(\tilde s,s')\bigr)V^i(s')\right]+\gamma\sum_{s'}T(s,s')\frac{\partial V^i(s')}{\partial z^i(\tilde s,\tilde a)} \\
        &=\mathbf1_{\{s=\tilde s\}}X^i(\tilde s,\tilde a)[Q^i(\tilde s,\tilde a)-V^i(\tilde s)]
        +\gamma\sum_{s'}T(s,s')\frac{\partial V^i(s')}{\partial z^i(\tilde s,\tilde a)} \\
    \end{aligned}
\end{equation*}
Consequently, we have
\begin{equation}
    \frac{\partial V^i(s)}{\partial z^i(\tilde s,\tilde a)}
        -\gamma\sum_{s'}T(s,s')\frac{\partial V^i(s')}{\partial z^i(\tilde s,\tilde a)}
        =\mathbf1_{\{s=\tilde s\}}X^i(\tilde s,\tilde a)
        \left[Q^i(\tilde s,\tilde a)-V^i(\tilde s)\right].
    \label{eq:si-value-derivative-components}
\end{equation}

We now collect these componentwise expressions into matrix and vector form. Let $\bm{e}_{\tilde s}$ denote the $K$-dimensional unit column with its only nonzero entry at state $\tilde s$, and let $\bm{T}^i(\tilde s,\tilde a,\cdot)$ and $\bm{T}(\tilde s,\cdot)$ denote transition row vectors. Equation~\eqref{eq:si-transition-reward-derivatives} gives a matrix with only row $\tilde s$ nonzero, while Eq.~\eqref{eq:si-value-derivative-components} is a linear system with coefficient matrix $\bm{I}-\gamma\bm T$. Since this matrix is invertible,
\begin{equation}
    \begin{aligned}
        \frac{\partial\bm T}{\partial z^i(\tilde s,\tilde a)}
        &=\bm{e}_{\tilde s}X^i(\tilde s,\tilde a)\left[\bm{T}^i(\tilde s,\tilde a,\cdot)-\bm{T}(\tilde s,\cdot)\right],\\
        \frac{\partial\bm V^i}{\partial z^i(\tilde s,\tilde a)}
        &=(\bm{I}-\gamma\bm T)^{-1}\bm{e}_{\tilde s}X^i(\tilde s,\tilde a)
        \left[Q^i(\tilde s,\tilde a)-V^i(\tilde s)\right].
    \end{aligned}
    \label{eq:si-bellman-derivatives}
\end{equation}

To compute the divergence, we first differentiate each action gap with respect to its own log-ratio coordinate. Recalling that $D^i(s,a)=Q^i(s,a)-Q^i(s,a_M)$ and that the conditional rewards and transition probabilities are independent of player $i$'s policy, we obtain
\begin{equation}
    \begin{aligned}
        \frac{\partial D^i(s,a)}{\partial z^i(s,a)}
        &=\gamma\sum_{s'}\left[T^i(s,a,s')-T^i(s,a_M,s')\right]
        \frac{\partial V^i(s')}{\partial z^i(s,a)}\\
        &=\gamma\left[\bm{T}^i(s,a,\cdot)-\bm{T}^i(s,a_M,\cdot)\right]
        \frac{\partial\bm V^i}{\partial z^i(s,a)}\\
        &=\gamma X^i(s,a)\left[Q^i(s,a)-V^i(s)\right] \left[\bm{T}^i(s,a,\cdot)-\bm{T}^i(s,a_M,\cdot)\right]
        (\bm{I}-\gamma\bm T)^{-1}\bm{e}_s.
    \end{aligned}
    \label{eq:si-gap-diagonal-derivative}
\end{equation}

We next compute the derivative of $\log w$. Since $w>0$, Jacobi's determinant formula gives, for a general coordinate,
\begin{equation*}
    \frac{\partial\log w}{\partial z^i(\tilde s,\tilde a)}
    =-\gamma\operatorname{tr}\left[
    (\bm{I}-\gamma\bm T)^{-1}\frac{\partial\bm T}{\partial z^i(\tilde s,\tilde a)}\right].
\end{equation*}
Evaluating at the same coordinate $z^i(s,a)$ as above, we substitute the transition derivative from Eq.~\eqref{eq:si-bellman-derivatives} and factor the scalar $X^i(s,a)$ out of the trace. Using $\operatorname{tr}(\bm{u}\bm{v})=\bm{v}\bm{u}$ for a column $\bm{u}$ and a row $\bm{v}$, so that the trace of their outer product equals their scalar product in the reverse order, we obtain
\begin{equation}
    \begin{aligned}
        \frac{\partial\log w}{\partial z^i(s,a)}
        &=-\gamma X^i(s,a)\operatorname{tr}\Big[
        \underbrace{(\bm{I}-\gamma\bm T)^{-1}\bm{e}_s}_{K\times1}
        \underbrace{\left[\bm{T}^i(s,a,\cdot)-\bm{T}(s,\cdot)\right]}_{1\times K}
        \Big]\\
        &=-\gamma X^i(s,a)\left[\bm{T}^i(s,a,\cdot)-\bm{T}(s,\cdot)\right]
        (\bm{I}-\gamma\bm T)^{-1}\bm{e}_s.
    \end{aligned}
    \label{eq:si-log-density-derivative}
\end{equation}

These two expressions cancel after summing over the actions with the appropriate action-gap factors. To see this explicitly, first note that
\begin{equation*}
    \sum_{a}X^i(s,a)\left[Q^i(s,a)-V^i(s)\right]=0,
    \qquad
    \sum_{a}X^i(s,a)\left[\bm{T}^i(s,a,\cdot)-\bm{T}(s,\cdot)\right]=0.
\end{equation*}
The reference-action term is zero in each sum below, so we may sum over all actions. Expanding and using these zero-mean identities gives
\begin{equation}
    \begin{aligned}
        &\sum_{a\ne a_M}X^i(s,a)
        \left[Q^i(s,a)-V^i(s)\right]
        \left[\bm{T}^i(s,a,\cdot)-\bm{T}^i(s,a_M,\cdot)\right] \\
        &\quad=\sum_{a\in\mathcal A}X^i(s,a)
        \left[Q^i(s,a)-V^i(s)\right]
        \left[\bm{T}^i(s,a,\cdot)-\bm{T}^i(s,a_M,\cdot)\right]\\
        &\quad=\sum_{a\in\mathcal A}X^i(s,a)
        \left[Q^i(s,a)-V^i(s)\right]\bm{T}^i(s,a,\cdot)\\
        &\quad=\sum_{a\in\mathcal A}X^i(s,a)Q^i(s,a)\bm{T}^i(s,a,\cdot)
        -V^i(s)\bm{T}(s,\cdot)\\
        &\quad=\sum_{a\in\mathcal A}X^i(s,a)Q^i(s,a)
        \left[\bm{T}^i(s,a,\cdot)-\bm{T}(s,\cdot)\right]\\
        &\quad=\sum_{a\ne a_M}X^i(s,a)D^i(s,a)
        \left[\bm{T}^i(s,a,\cdot)-\bm{T}(s,\cdot)\right].
    \end{aligned}
    \label{eq:si-covariance-identity}
\end{equation}
In the last step, subtracting $Q^i(s,a_M)$ from every action value does not change the sum. Combining this identity with Eqs.~\eqref{eq:si-gap-diagonal-derivative} and~\eqref{eq:si-log-density-derivative}, and summing over all players and states, yields
\begin{equation*}
    \operatorname{div}\bm D
    =\sum_{i,s}\sum_{a\ne a_M}\frac{\partial D^i(s,a)}{\partial z^i(s,a)}
    =-\sum_{i,s}\sum_{a\ne a_M}D^i(s,a)
    \frac{\partial\log w}{\partial z^i(s,a)}
    =-\bm D\cdot\nabla\log w.
\end{equation*}
The product rule now proves the first claim:
\begin{equation*}
    \operatorname{div}(w\bm D)
    =w\operatorname{div}\bm D+\bm D\cdot\nabla w
    =w\left(\operatorname{div}\bm D+\bm D\cdot\nabla\log w\right)=0.
\end{equation*}

Finally, we show how this local identity implies conservation of the weighted volume of an evolving region. For an initial point $\bm{z}_0$, write $\bm{z}(t)=\Phi_t(\bm{z}_0)$. The smooth flow is invertible, and Liouville's formula for its Jacobian determinant gives
\begin{equation*}
    \frac{\mathrm{d}}{\mathrm{d}t}
    \log\det\left(\frac{\partial\Phi_t(\bm{z}_0)}{\partial \bm{z}_0}\right)
    =\operatorname{div}\bm D(\Phi_t(\bm{z}_0)).
\end{equation*}
The determinant starts at one and stays positive. Along the same trajectory, the chain rule gives
$\mathrm{d}\log w(\Phi_t(\bm{z}_0))/\mathrm{d}t
=\bm D(\Phi_t(\bm{z}_0))\cdot\nabla\log w(\Phi_t(\bm{z}_0))$.
Their sum is zero by the divergence identity just proved. Integrating in time therefore yields
\begin{equation*}
    w(\Phi_t(\bm{z}_0))
    \det\left(\frac{\partial\Phi_t(\bm{z}_0)}{\partial \bm{z}_0}\right)
    =w(\bm{z}_0).
\end{equation*}
Changing variables from $\bm{z}\in U_t$ to $\bm{z}_0\in U_0$ then gives
\begin{equation*}
    \int_{U_t}w(\bm{z})\,\mathrm{d}\bm{z}
    =\int_{U_0}w(\Phi_t(\bm{z}_0))
    \det\left(\frac{\partial\Phi_t(\bm{z}_0)}{\partial \bm{z}_0}\right)\,\mathrm{d}\bm{z}_0
    =\int_{U_0}w(\bm{z}_0)\,\mathrm{d}\bm{z}_0,
\end{equation*}
which proves Eq.~\eqref{eq:si-weighted-volume-integral}.
\end{proof}

Volume preservation also excludes an asymptotically stable fully interior equilibrium. Otherwise, a compact set of positive weighted volume in its attraction basin would eventually enter arbitrarily small neighborhoods of that equilibrium, contradicting Eq.~\eqref{eq:si-weighted-volume-integral} and the upper bound in Lemma~\ref{lem:si-volume-weight}. Boundary attraction remains possible, since boundary policies lie at infinite log-ratio distance.

\subsection{Dissipation in the presence of exploration}

With entropy regularization, write $\bm{J}(\bm{z})=\partial\bm D(\bm{z})/\partial \bm{z}$ for the Jacobian of the action-value gaps. At a fixed policy point,
\begin{equation}
    \frac{\partial}{\partial \bm{z}}\left[\bm D(\bm{z})-\eta \bm{z}\right]=\bm{J}(\bm{z})-\eta \bm{I},
    \qquad
    \operatorname{div}(\bm D-\eta \bm{z})=\operatorname{div}\bm D-d\eta.
    \label{eq:si-entropy-jacobian}
\end{equation}
Entropy therefore adds uniform local damping and reduces the ordinary divergence by $d\eta$ at the same policy point.

\subsection{Interior boundedness and equilibrium existence}
\label{sec:si-interior-existence}

Exploration prevents action preferences from growing without bound. In Eq.~\eqref{eq:si-reduced-log-ode}, the value gaps $\bm D(\bm z)$ that strengthen preferences are bounded by $B$ in Eq.~\eqref{eq:si-gap-bound}, whereas the restoring term $-\eta\bm z$ grows linearly with preference strength. Once a log ratio exceeds $B/\eta$ in magnitude, the restoring term dominates and pulls it back. The following lemma makes this precise. We use it to locate the equilibria in Proposition~\ref{prop:si-equilibrium-existence}.

\begin{lemma}[Exploration confines learning to the interior]
\label{lem:si-interior-boundedness}
Let $\eta>0$ and define the box $\mathcal K_\eta=\{\bm z\in\mathbb R^d:\|\bm z\|_\infty\leq B/\eta\}$. Every solution of Eq.~\eqref{eq:si-reduced-log-ode} satisfies, for each coordinate,
\begin{equation}
    |z^i(s,a;t)|-\frac{B}{\eta}
    \leq e^{-\eta t}\left(|z^i(s,a;0)|-\frac{B}{\eta}\right),
    \qquad t\geq0.
    \label{eq:si-log-bound}
\end{equation}
Hence $\mathcal K_\eta$ is forward invariant, and every trajectory starting outside $\mathcal K_\eta$ approaches it exponentially at rate $\eta$. Moreover, every policy profile in $\mathcal K_\eta$ assigns probability at least
\begin{equation*}
    p_\eta=\frac{e^{-B/\eta}}{1+(M-1)e^{B/\eta}}>0
\end{equation*}
to every action in every state.
\end{lemma}

\begin{proof}
Fix a coordinate $z^i(s,a)$. Since $\mathrm{d}(e^{\eta t}z^i(s,a))/\mathrm{d}t=e^{\eta t}D^i(s,a)$, we have
\begin{equation*}
    z^i(s,a;t)=e^{-\eta t}z^i(s,a;0)+\int_0^te^{-\eta(t-u)}D^i\bigl(s,a;\bm z(u)\bigr)\,\mathrm{d}u.
\end{equation*}
Using $|D^i(s,a)|\leq B$ from Eq.~\eqref{eq:si-gap-bound}, we obtain
\begin{equation*}
    |z^i(s,a;t)|\leq e^{-\eta t}|z^i(s,a;0)|+\frac{B}{\eta}\left(1-e^{-\eta t}\right),
\end{equation*}
which rearranges to Eq.~\eqref{eq:si-log-bound}. If $\bm z(0)\in\mathcal K_\eta$, the right-hand side of Eq.~\eqref{eq:si-log-bound} is nonpositive, so $\bm z(t)\in\mathcal K_\eta$ for all $t\geq0$. For the probability bound, let $\|\bm z\|_\infty\leq B/\eta$. In Eq.~\eqref{eq:si-log-inverse}, each numerator is at least $e^{-B/\eta}$, since $z^i(s,a)\geq-B/\eta$ and the reference action has numerator $1$, while the denominator is at most $1+(M-1)e^{B/\eta}$. Hence $X^i(s,a)\geq p_\eta$ for every action.
\end{proof}

\begin{proposition}
\label{prop:si-equilibrium-existence}
For every $\eta>0$, Eq.~\eqref{eq:si-reduced-log-ode} has at least one equilibrium $\bm z^{\ast}\in\mathcal K_\eta$. The corresponding policy profile $\bm X^{\ast}$ is fully interior, assigns probability at least $p_\eta$ to every action, and satisfies
\begin{equation}
    X^{i\ast}(s,a)=\frac{\exp\bigl(Q^i(s,a)/\eta\bigr)}{\sum_{b\in\mathcal A}\exp\bigl(Q^i(s,b)/\eta\bigr)},
    \label{eq:si-equilibrium-map}
\end{equation}
where the Bellman values are evaluated at $\bm X^{\ast}$.
\end{proposition}

\begin{proof}
An equilibrium of Eq.~\eqref{eq:si-reduced-log-ode} satisfies $\bm D(\bm z^{\ast})=\eta\bm z^{\ast}$, so it is a fixed point of the map $\bm G(\bm z)=\bm D(\bm z)/\eta$. This map is continuous because $\bm D$ is smooth. By Eq.~\eqref{eq:si-gap-bound}, $\|\bm G(\bm z)\|_\infty\leq B/\eta$ for every $\bm z$, so $\bm G$ maps the compact convex set $\mathcal K_\eta$ into itself. Brouwer's fixed-point theorem gives a fixed point $\bm z^{\ast}\in\mathcal K_\eta$. By Lemma~\ref{lem:si-interior-boundedness}, every action has probability at least $p_\eta$ under $\bm X^{\ast}$. Finally, $z^{i\ast}(s,a)=D^i(s,a)/\eta$ means $X^{i\ast}(s,a)/X^{i\ast}(s,a_M)=\exp\{[Q^i(s,a)-Q^i(s,a_M)]/\eta\}$, and normalizing over actions gives Eq.~\eqref{eq:si-equilibrium-map}.
\end{proof}

\subsection{Proof of Proposition~1}

\begin{mainproposition}
\label{prop:si-exploration-local-stability}
An interior equilibrium $\bm z^{\ast}$ is exponentially stable if and only if
\begin{equation}
    \max_k \operatorname{Re} \lambda_k(\bm{J}^{\ast}) < \eta,
    \label{eq:si-exploration-local-stability}
\end{equation}
where $\lambda_k(\bm{J}^{\ast})$ are the eigenvalues of $\bm{J}^{\ast}$.
\end{mainproposition}

\begin{proof}
Write $\bm{J}^{\ast}=\partial\bm D(\bm z^{\ast})/\partial\bm z$ for the Jacobian
of value feedback at the equilibrium. The Jacobian of the full vector
field $\bm D(\bm z)-\eta\bm z$ at $\bm z^{\ast}$ is
$\bm{J}^{\ast}-\eta\bm I$ and its eigenvalues are $\lambda_k(\bm{J}^{\ast})-\eta$. The equilibrium is exponentially stable if and only if all eigenvalues of its Jacobian have strictly negative real parts.
\end{proof}

\clearpage

\bibliographystyle{unsrtnat}
\bibliography{ref}

@book{russell2010artificial,
  title={Artificial Intelligence: A Modern Approach},
  author={Russell, Stuart and Norvig, Peter},
  year={2010},
  publisher={Prentice Hall},
  edition={3rd},
  address={Upper Saddle River, NJ}
}

@book{sutton1998reinforcement,
  title={Reinforcement Learning: An Introduction},
  author={Sutton, Richard S and Barto, Andrew G},
  year={1998},
  publisher = {MIT Press},
  address = {Cambridge, MA}
}

@article{fawzi2022discovering,
  title={Discovering faster matrix multiplication algorithms with reinforcement learning},
  author={Fawzi, Alhussein and Balog, Matej and Huang, Aja and Hubert, Thomas and Romera-Paredes, Bernardino and Barekatain, Mohammadamin and Novikov, Alexander and Ruiz, Francisco J. R. and Schrittwieser, Julian and Swirszcz, Grzegorz and others},
  journal={Nature},
  volume={610},
  number={7930},
  pages={47--53},
  year={2022},
  publisher={Nature Publishing Group UK, London}
}

@article{lee2004reinforcement,
  title={Reinforcement learning and decision making in monkeys during a competitive game},
  author={Lee, Daeyeol and Conroy, Michelle L and McGreevy, Benjamin P and Barraclough, Dominic J},
  journal={Cogn Brain Res},
  volume={22},
  number={1},
  pages={45--58},
  year={2004},
  publisher={Elsevier}
}

@article{sgroi2021reinforcement,
  title={Reinforcement learning approach to nonequilibrium quantum thermodynamics},
  author={Sgroi, Sofia and Palma, G Massimo and Paternostro, Mauro},
  journal={Phys Rev Lett},
  volume={126},
  number={2},
  pages={020601},
  year={2021},
  publisher={APS}
}

@article{silver2017mastering,
  title={Mastering the game of {Go} without human knowledge},
  author={Silver, David and Schrittwieser, Julian and Simonyan, Karen and Antonoglou, Ioannis and Huang, Aja and Guez, Arthur and Hubert, Thomas and Baker, Lucas and Lai, Matthew and Bolton, Adrian and others},
  journal={Nature},
  volume={550},
  number={7676},
  pages={354--359},
  year={2017},
  publisher={Nature Publishing Group UK London}
}

@inproceedings{ouyang2022training,
  title={Training language models to follow instructions with human feedback},
  author={Ouyang, Long and Wu, Jeffrey and Jiang, Xu and Almeida, Diogo and Wainwright, Carroll and Mishkin, Pamela and Zhang, Chong and Agarwal, Sandhini and Slama, Katarina and Ray, Alex and others},
  booktitle = {Advances in Neural Information Processing Systems},
  volume={35},
  pages={27730--27744},
  year={2022},
  editor={Koyejo, S. and Mohamed, S. and Agarwal, A. and Belgrave, D. and Cho, K. and Oh, A.},
  publisher={Curran Associates, Inc.}
}

@article{geng2026learning,
  title={Learning to cooperate in a changing world: How caring about the future promotes cooperation across scales},
  author={Geng, Yuxin and Chen, Xingru and Wang, Xin and Zheng, Hongwei and Liu, Longzhao and Tang, Shaoting and Fu, Feng},
  journal={arXiv preprint arXiv:2609.34005},
  year={2026}
}

@article{sutton1988learning,
  title={Learning to predict by the methods of temporal differences},
  author={Sutton, Richard S},
  journal={Mach Learn},
  volume={3},
  number={1},
  pages={9--44},
  year={1988},
  publisher={Springer}
}

@article{watkins1992q,
  title={{Q}-learning},
  author={Watkins, Christopher J C H and Dayan, Peter},
  journal={Mach Learn},
  volume={8},
  number={3--4},
  pages={279--292},
  year={1992},
  publisher={Springer}
}

@inproceedings{konda1999actor,
  title={Actor-critic algorithms},
  author={Konda, Vijay R. and Tsitsiklis, John N.},
  booktitle={Advances in Neural Information Processing Systems},
  volume={12},
  year={1999},
  pages={1008--1014},
  editor={Solla, S. and Leen, T. and M{\"u}ller, K.},
  publisher={MIT Press},
  address={Cambridge, MA}
}

@article{busoniu2008comprehensive,
  title={A comprehensive survey of multiagent reinforcement learning},
  author={Busoniu, Lucian and Babuska, Robert and De Schutter, Bart},
  journal={IEEE Trans Syst Man Cybern C Appl Rev},
  volume={38},
  number={2},
  pages={156--172},
  year={2008},
  publisher={IEEE}
}

@article{tuyls2005evolutionary,
  title={Evolutionary game theory and multi-agent reinforcement learning},
  author={Tuyls, Karl and Now{\'e}, Ann},
  journal={Knowl Eng Rev},
  volume={20},
  number={1},
  pages={63--90},
  year={2005},
  publisher={Cambridge University Press}
}

@book{wiering2012reinforcement,
  editor    = {Wiering, Marco and van Otterlo, Martijn},
  title     = {Reinforcement Learning: State-of-the-Art},
  address   = {Berlin, Heidelberg},
  publisher = {Springer},
  year      = {2012}
}

@inproceedings{leibo2017multi,
  title = {Multi-agent reinforcement learning in sequential social dilemmas},
  author = {Leibo, Joel Z. and Zambaldi, Vinicius and Lanctot, Marc and Marecki, Janusz and Graepel, Thore},
  year = {2017},
  booktitle = {the 16th International Conference on Autonomous Agents and Multiagent Systems},
  pages = {464--473},
}

@inproceedings{jaques2019social,
  title={Social influence as intrinsic motivation for multi-agent deep reinforcement learning},
  author={Jaques, Natasha and Lazaridou, Angeliki and Hughes, Edward and Gulcehre, Caglar and Ortega, Pedro and Strouse, D J and Leibo, Joel Z and De Freitas, Nando},
  booktitle={the 36th International Conference on Machine Learning},
  pages={3040--3049},
  year={2019},
  volume={97},
  series={Proceedings of Machine Learning Research},
  editor={Chaudhuri, Kamalika and Salakhutdinov, Ruslan},
  publisher={PMLR}
}

@inproceedings{foerster2016learning,
  title={Learning to communicate with deep multi-agent reinforcement learning},
  author={Foerster, Jakob and Assael, Ioannis Alexandros and De Freitas, Nando and Whiteson, Shimon},
  booktitle={Advances in Neural Information Processing Systems},
  volume={29},
  year={2016},
  editor={Lee, D. and Sugiyama, M. and von Luxburg, U. and Guyon, I. and Garnett, R.},
  publisher={Curran Associates, Inc.}
}

@inproceedings{ndousse2021emergent,
  title={Emergent social learning via multi-agent reinforcement learning},
  author={Ndousse, Kamal K and Eck, Douglas and Levine, Sergey and Jaques, Natasha},
  booktitle={the 38th International Conference on Machine Learning},
  pages={7991--8004},
  year={2021},
  volume={139},
  series={Proceedings of Machine Learning Research},
  editor={Meila, Marina and Zhang, Tong},
  publisher={PMLR}
}

@article{yang2020overview,
  title={An overview of multi-agent reinforcement learning from game theoretical perspective},
  author={Yang, Yaodong and Wang, Jun},
  journal={arXiv preprint arXiv:2011.00583v1},
  year={2020}
}

@InProceedings{mnih2016asynchronous,
  title={Asynchronous methods for deep reinforcement learning},
  author={Mnih, Volodymyr and Badia, Adria Puigdomenech and Mirza, Mehdi and Graves, Alex and Lillicrap, Timothy and Harley, Tim and Silver, David and Kavukcuoglu, Koray},
  booktitle={the 33rd International Conference on Machine Learning},
  volume={48},
  pages={1928--1937},
  year={2016},
  series={Proceedings of Machine Learning Research},
  editor={Balcan, Maria Florina and Weinberger, Kilian Q.},
  publisher={PMLR},
  address={New York, New York, USA}
}

@article{haarnoja2018soft,
  title={Soft actor-critic algorithms and applications},
  author={Haarnoja, Tuomas and Zhou, Aurick and Hartikainen, Kristian and Tucker, George and Ha, Sehoon and Tan, Jie and Kumar, Vikash and Zhu, Henry and Gupta, Abhishek and Abbeel, Pieter and others},
  journal={arXiv preprint arXiv:1812.05905},
  year={2018}
}

@inproceedings{ahmed2019understanding,
  title={Understanding the impact of entropy on policy optimization},
  author={Ahmed, Zafarali and Le Roux, Nicolas and Norouzi, Mohammad and Schuurmans, Dale},
  booktitle={the 36th International Conference on Machine Learning},
  pages={151--160},
  year={2019},
  volume={97},
  series={Proceedings of Machine Learning Research},
  editor={Chaudhuri, Kamalika and Salakhutdinov, Ruslan},
  publisher={PMLR}
}

@article{barfuss2025collective,
  author = {Barfuss, Wolfram and Flack, Jessica and Gokhale, Chaitanya S. and Hammond, Lewis and Hilbe, Christian and Hughes, Edward and Leibo, Joel Z. and Lenaerts, Tom and Leonard, Naomi and Levin, Simon and Madhushani Sehwag, Udari and McAvoy, Alex and Meylahn, Janusz M. and Santos, Fernando P.},
  title = {Collective cooperative intelligence},
  journal = {Proc Natl Acad Sci USA},
  volume = {122},
  number = {25},
  pages = {e2319948121},
  year = {2025}
}

@article{schulman2017proximal,
  title={Proximal policy optimization algorithms},
  author={Schulman, John and Wolski, Filip and Dhariwal, Prafulla and Radford, Alec and Klimov, Oleg},
  journal={arXiv preprint arXiv:1707.06347},
  year={2017}
}

@article{shapley1953stochastic,
  title={Stochastic games},
  author={Shapley, Lloyd S},
  journal={Proc Natl Acad Sci USA},
  volume={39},
  number={10},
  pages={1095--1100},
  year={1953},
  publisher={National Academy of Sciences}
}

@article{hilbe2018evolution,
  title={Evolution of cooperation in stochastic games},
  author={Hilbe, Christian and {\v{S}}imsa, {\v{S}}t{\v{e}}p{\'a}n and Chatterjee, Krishnendu and Nowak, Martin A},
  journal={Nature},
  volume={559},
  number={7713},
  pages={246--249},
  year={2018},
  publisher={Nature Publishing Group UK London}
}

@article{borgers1997learning,
  title={Learning through reinforcement and replicator dynamics},
  author={B{\"o}rgers, Tilman and Sarin, Rajiv},
  journal={J Econ Theory},
  volume={77},
  number={1},
  pages={1--14},
  year={1997},
  publisher={Elsevier}
}

@article{wang2024mathematics,
  title={Mathematics of multi-agent learning systems at the interface of game theory and artificial intelligence},
  author={Wang, Long and Fu, Feng and Chen, Xingru},
  journal={Sci China Inf Sci},
  volume={67},
  number={6},
  year={2024},
  publisher={Science China Press},
  pages={166201}
}

@book{nowak2006evolutionary,
  title={Evolutionary Dynamics: Exploring the Equations of Life},
  author={Nowak, Martin A},
  year={2006},
  publisher={Belknap Press of Harvard University Press},
  address={Cambridge, MA}
}

@inproceedings{tuyls2003selection,
  title={A selection-mutation model for {Q}-learning in multi-agent systems},
  author={Tuyls, Karl and Verbeeck, Katja and Lenaerts, Tom},
  booktitle={the 2nd International Joint Conference on Autonomous Agents and Multiagent Systems},
  pages={693--700},
  year={2003}
}

@article{kianercy2012dynamics,
  title={Dynamics of {Boltzmann} {$Q$} learning in two-player two-action games},
  author={Kianercy, Ardeshir and Galstyan, Aram},
  journal={Phys Rev E},
  volume={85},
  number={4},
  pages={041145},
  year={2012},
  publisher={APS}
}

@inproceedings{kaisers2010frequency,
  title={Frequency adjusted multi-agent {Q}-learning},
  author={Kaisers, Michael and Tuyls, Karl},
  booktitle={the 9th International Conference on Autonomous Agents and Multiagent Systems},
  pages={309--315},
  year={2010}
}

@article{leonardos2022exploration,
  title={Exploration-exploitation in multi-agent learning: Catastrophe theory meets game theory},
  author={Leonardos, Stefanos and Piliouras, Georgios},
  journal={Artif Intell},
  volume={304},
  pages={103653},
  year={2022},
  publisher={Elsevier}
}

@inproceedings{littman1994markov,
  title={{Markov} games as a framework for multi-agent reinforcement learning},
  author={Littman, Michael L},
  booktitle={the 11th International Conference on Machine Learning},
  pages={157--163},
  year={1994},
}

@inproceedings{hennes2009state,
author = {Hennes, Daniel and Tuyls, Karl and Rauterberg, Matthias},
title = {State-coupled replicator dynamics},
booktitle = {the 8th International Conference on Autonomous Agents and Multiagent Systems},
year = {2009},

pages = {789--796},
}

@inproceedings{hennes2010resq,
  title={{RESQ}-learning in stochastic games},
  author={Hennes, Daniel and Kaisers, Michael and Tuyls, Karl},
  booktitle={Adaptive and Learning Agents Workshop at AAMAS},
  year={2010},
  pages={8--15}
}

@article{barfuss2019deterministic,
  title={Deterministic limit of temporal difference reinforcement learning for stochastic games},
  author={Barfuss, Wolfram and Donges, Jonathan F and Kurths, J{\"u}rgen},
  journal={Phys Rev E},
  volume={99},
  number={4},
  pages={043305},
  year={2019},
  publisher={APS}
}

@inproceedings{geng2026emergent,
  title={Emergent fast-slow dynamics in multi-agent {Q}-learning for networked stochastic games},
  author={Geng, Yuxin and Barfuss, Wolfram and Chen, Xingru},
  booktitle={the 40th AAAI Conference on Artificial Intelligence},
  volume={40},
  number={35},
  pages={29450--29458},
  year={2026}
}

@article{beck2003mirror,
  title={Mirror descent and nonlinear projected subgradient methods for convex optimization},
  author={Beck, Amir and Teboulle, Marc},
  journal={Oper Res Lett},
  volume={31},
  number={3},
  pages={167--175},
  year={2003},
  publisher={Elsevier}
}

@book{souchay2010dynamics,
  title={Dynamics of Small Solar System Bodies and Exoplanets},
  editor={Souchay, Jean J. and Dvorak, Rudolf},
  series={Lecture Notes in Physics},
  volume={790},
  year={2010},
  publisher={Springer},
  address={Berlin, Heidelberg},
  doi={10.1007/978-3-642-04458-8}
}

@article{nowak2006five,
  title={Five rules for the evolution of cooperation},
  author={Nowak, Martin A},
  journal={Science},
  volume={314},
  number={5805},
  pages={1560--1563},
  year={2006},
  publisher={American Association for the Advancement of Science}
}

@article{glynatsi2024conditional,
  title={Conditional cooperation with longer memory},
  author={Glynatsi, Nikoleta E and Akin, Ethan and Nowak, Martin A and Hilbe, Christian},
  journal={Proc Natl Acad Sci USA},
  volume={121},
  number={50},
  pages={e2420125121},
  year={2024},
  publisher={National Academy of Sciences}
}

@article{nowak1998evolution,
  title={Evolution of indirect reciprocity by image scoring},
  author={Nowak, Martin A and Sigmund, Karl},
  journal={Nature},
  volume={393},
  number={6685},
  pages={573--577},
  year={1998},
  publisher={Nature Publishing Group UK, London}
}

@article{ohtsuki2006leading,
  title={The leading eight: social norms that can maintain cooperation by indirect reciprocity},
  author={Ohtsuki, Hisashi and Iwasa, Yoh},
  journal={J Theor Biol},
  volume={239},
  number={4},
  pages={435--444},
  year={2006},
  publisher={Elsevier}
}

@article{santos2018social,
  title={Social norm complexity and past reputations in the evolution of cooperation},
  author={Santos, Fernando P and Santos, Francisco C and Pacheco, Jorge M},
  journal={Nature},
  volume={555},
  number={7695},
  pages={242--245},
  year={2018},
  publisher={Nature Publishing Group UK, London}
}

@article{sutton1999between,
  title={Between {MDPs} and semi-{MDPs}: A framework for temporal abstraction in reinforcement learning},
  author={Sutton, Richard S and Precup, Doina and Singh, Satinder},
  journal={Artif Intell},
  volume={112},
  number={1--2},
  pages={181--211},
  year={1999},
  publisher={Elsevier}
}

@article{chen2023outlearning,
  title={Outlearning extortioners: unbending strategies can foster reciprocal fairness and cooperation},
  author={Chen, Xingru and Fu, Feng},
  journal={PNAS Nexus},
  volume={2},
  number={6},
  pages={pgad176},
  year={2023},
  publisher={Oxford University Press, US}
}

@article{pal2026large,
  title={Large language models instantiate evolutionarily robust strategies of cooperation},
  author={Pal, Saptarshi and Mallela, Abhishek and Pracher, Lenz and Wei, Chiyu and Fu, Feng and Schnell, Santiago and Nowak, Martin A},
  journal={PNAS Nexus},
  volume={5},
  number={6},
  pages={pgag210},
  year={2026},
  publisher={Oxford University Press, US}
}

@article{fu2026optimal,
  title={On the optimal integration of intelligent agents into network systems to steer cooperation},
  author={Fu, Feng and Chen, Xingru and Christakis, Nicholas A},
  journal={Proc Natl Acad Sci USA},
  volume={123},
  number={12},
  pages={e2537939123},
  year={2026},
  publisher={National Academy of Sciences}
}

@article{liu2023coevolutionary,
  title={Coevolutionary dynamics via adaptive feedback in collective-risk social dilemma game},
  author={Liu, Linjie and Chen, Xiaojie and Szolnoki, Attila},
  journal={eLife},
  volume={12},
  pages={e82954},
  year={2023},
  publisher={eLife Sciences Publications, Ltd}
}

@article{chen2015competition,
  title={Competition and cooperation among different punishing strategies in the spatial public goods game},
  author={Chen, Xiaojie and Szolnoki, Attila and Perc, Matja{\v{z}}},
  journal={Phys Rev E},
  volume={92},
  number={1},
  pages={012819},
  year={2015},
  publisher={APS}
}

@article{zhang2014evolution,
  title={The evolution of sanctioning institutions: an experimental approach to the social contract},
  author={Zhang, Boyu and Li, Cong and De Silva, Hannelore and Bednarik, Peter and Sigmund, Karl},
  journal={Exp Econ},
  volume={17},
  number={2},
  pages={285--303},
  year={2014},
  publisher={Springer}
}

@article{du2014aspiration,
  title={Aspiration dynamics of multi-player games in finite populations},
  author={Du, Jinming and Wu, Bin and Altrock, Philipp M and Wang, Long},
  journal={J R Soc Interface},
  volume={11},
  number={94},
  pages={20140077},
  year={2014}
}

@article{cao2019dynamic,
  title={Dynamic matching pennies on networks},
  author={Cao, Zhigang and Qin, Cheng-Zhong and Yang, Xiaoguang and Zhang, Boyu},
  journal={Int J Game Theory},
  volume={48},
  number={3},
  pages={887--920},
  year={2019},
  publisher={Springer}
}

@article{rong2010feedback,
  title={Feedback reciprocity mechanism promotes the cooperation of highly clustered scale-free networks},
  author={Rong, Zhihai and Yang, Han-Xin and Wang, Wen-Xu},
  journal={Phys Rev E},
  volume={82},
  number={4},
  pages={047101},
  year={2010},
  publisher={APS}
}

@article{allen2017evolutionary,
  title={Evolutionary dynamics on any population structure},
  author={Allen, Benjamin and Lippner, Gabor and Chen, Yu-Ting and Fotouhi, Babak and Momeni, Naghmeh and Yau, Shing-Tung and Nowak, Martin A},
  journal={Nature},
  volume={544},
  number={7649},
  pages={227--230},
  year={2017},
  publisher={Nature Publishing Group UK, London}
}

\end{document}